\documentclass[compsoc,conference,a4paper,10pt,times]{IEEEtran}
\usepackage[T1]{fontenc}
\usepackage{cite}
\usepackage{array,booktabs,tabularx,adjustbox}
\usepackage[table]{xcolor}
\usepackage{float,graphicx}
\usepackage{textcomp,amssymb,amsmath,amsthm}
\usepackage{url,xspace,multirow,enumitem,pifont}
\usepackage{tikz}
\usetikzlibrary{positioning,arrows.meta}
\newtheorem{remark}{Remark}
\newtheorem{definition}{Definition}
\newtheorem{theorem}{Theorem}
\newtheorem{lemma}{Lemma}
\renewcommand{\emph}[1]{\textit{#1}}
\definecolor{rowalt}{gray}{0.96}
\newcommand{\cmark}{\ding{51}}%
\newcommand{\xmark}{\ding{55}}%
\newcommand{\RootAnchor}{\ensuremath{\mathsf{RootAnchor}}}
\newcommand{\Fp}{\ensuremath{\mathbb{F}_p}\xspace}
\newcommand{\F}{\ensuremath{\mathsf{F}}\xspace}

\newcommand{\negl}{\ensuremath{\mathsf{negl}}\xspace}

\newcommand{\ipk}{\ensuremath{\mathsf{ipk}}\xspace}
\newcommand{\isk}{\ensuremath{\mathsf{isk}}\xspace}
\newcommand{\upk}{\ensuremath{\mathsf{upk}}\xspace}
\newcommand{\usk}{\ensuremath{\mathsf{usk}}\xspace}
\newcommand{\cred}{\ensuremath{\mathsf{Cred}}\xspace}
\newcommand{\credl}{\ensuremath{\mathsf{Cred}_L}\xspace}

\newcommand{\pcred}{\ensuremath{\mathsf{PCCred}}\xspace}
\newcommand{\setup}{\ensuremath{\mathsf{Setup}}\xspace}
\newcommand{\keygen}{\ensuremath{\mathsf{KeyGen}}\xspace}

\newcommand{\prove}{\ensuremath{\mathsf{Prove}}\xspace}
\newcommand{\cverify}{\ensuremath{\mathsf{Verify}}\xspace}
\newcommand{\PK}{\ensuremath{\mathsf{PK}}\xspace}
\newcommand{\relation}{\ensuremath{\mathsf{R}}\xspace}

\newcommand{\proverk}{\ensuremath{\mathsf{pk}}\xspace}
\newcommand{\verifierk}{\ensuremath{\mathsf{vk}}\xspace}
\newcommand{\pp}{\ensuremath{\mathsf{pp}}\xspace}

\newcommand{\sk}{\ensuremath{\mathsf{sk}}\xspace}
\newcommand{\pk}{\ensuremath{\mathsf{pk}}\xspace}
\newcommand{\nivcsetup}{\ensuremath{\mathsf{NIVC.Setup}}\xspace}

\newcommand{\nivckeygen}{\ensuremath{\mathsf{NIVC.KeyGen}}\xspace}

\newcommand{\nivcprove}{\ensuremath{\mathsf{NIVC.Prove}}\xspace}

\newcommand{\nivcverify}{\ensuremath{\mathsf{NIVC.Verify}}\xspace}

\newcommand{\signverify}{\ensuremath{\Sigma.\mathsf{Verify}}\xspace}

\newcommand{\pcc}{\ensuremath{\Pi_{\mathsf{PCC}}}\xspace}

\newcommand{\pccsetup}{\ensuremath{\mathsf{PCC.Setup}}\xspace}
\newcommand{\pcckeygen}{\ensuremath{\mathsf{PCC.KeyGen}}\xspace}
\newcommand{\pccderive}{\ensuremath{\mathsf{PCC.Derive}}\xspace}
\newcommand{\pccverify}{\ensuremath{\mathsf{PCC.Verify}}\xspace}
\newcommand{\token}{\ensuremath{\mathsf{Token}}\xspace}
\newcommand{\policy}{\ensuremath{\mathsf{Policy}}\xspace}

\newcommand{\commit}{\ensuremath{\mathsf{Com}}\xspace} 
 
\newcommand{\phash}{\ensuremath{\mathsf{H}}\xspace}

\newcommand{\spar}[1]{%
  \noindent%
  \textbf{#1}\xspace%
}

\newcommand{\researchquestion}[1]{%
  \par\vspace{0.15\baselineskip}
  \noindent
  \begin{minipage}{\linewidth}
    \centering\itshape
    #1
  \end{minipage}
  \par\vspace{0.15\baselineskip}
}
\usepackage{
  tikz,
  pgfplots,
  pgfplotstable
}
\usepackage{hyperref}

\usetikzlibrary{
  shapes.geometric,
  arrows,
  external,
  pgfplots.groupplots,
  matrix
}

\pgfplotsset{compat=1.9}
\usepackage{mathtools,}

\DeclareMathAlphabet{\mathcal}{OMS}{cmsy}{m}{n}
\DeclareGraphicsExtensions{%
    .png,.PNG,%
    .pdf,.PDF,%
    .jpg,.mps,.jpeg,.jbig2,.jb2,.JPG,.JPEG,.JBIG2,.JB2}

\hypersetup{colorlinks=true,urlcolor=black,linkcolor=black,citecolor=black}
\begin{document}
\title{PCC: Proof-Carrying Credentials for Enhanced Anonymity and Post-Quantum Delegation}
\author{
\IEEEauthorblockN{
Zixuan Wang\IEEEauthorrefmark{1},
ZHENG Tianyu\IEEEauthorrefmark{2},
Yuan Lu\IEEEauthorrefmark{3},
Bin Xiao\IEEEauthorrefmark{1}
}
\IEEEauthorblockA{
\IEEEauthorrefmark{1}The Hong Kong Polytechnic University\\
24052881r@connect.polyu.hk,\quad b.xiao@polyu.edu.hk
}
\IEEEauthorblockA{
\IEEEauthorrefmark{2}Department of Computing,
The Hong Kong Polytechnic University\\
tian-yu.zheng@connect.polyu.hk
}
\IEEEauthorblockA{
\IEEEauthorrefmark{3}Agency for Science, Technology and Research (A*STAR)\\
lu\_yuan@a-star.edu.sg
}
}
\maketitle
\begin{abstract}
Delegatable Anonymous Credentials (DACs) let an authority be delegated through a chain of intermediaries and later exercised anonymously, as in DeFi credit delegation.
Because presentation is the operation a holder repeats at every use, its cost is what matters in practice; yet existing DACs build this presentation from the delegation chain itself.
We call this limitation \emph{prover awareness}: to present, a holder must possess and process delegation-lineage material whose shape mirrors the chain.
It has two faces: its size makes the presentation cost grow with the delegation depth, and its content exposes the holder to delegation-topology metadata that, in incentive-driven and compliance-aware settings, enables lineage-based policy vetoes, strategic disintermediation, and de facto censorship in otherwise permissionless networks.

We formalize the privacy face as \emph{Enhanced Delegation Anonymity} (EDA), a leakage-parametrized two-observer notion combining verifier-side anonymity with holder-side topology obliviousness, the latter given both as a simulation definition and as an indistinguishability game over paths with identical allowed holder leakage.
Because both faces share one cause, the chain-structured credential, a single change of representation removes them together: our framework \emph{Proof-Carrying Credentials} (PCC) replaces the explicit chain with a single recursively updated proof-carrying state via Non-Uniform Incrementally Verifiable Computation (NIVC), making the presentation independent of the delegation depth and hiding the topology at once, while preserving restricted delegation and expressive presentation.
Unlike prior constant-size DACs tied to bilinear pairings, PCC is hardness-agnostic and admits an explicit post-quantum-compatible instantiation in the QROM from transparent FRI-based recursion and hash-based signatures.

On consumer-grade hardware, the Halo2 (non-post-quantum) instantiation keeps the verifier-facing proof constant at 7.344\,KB with 35--36\,ms verification across all tested depths, and a 128-bit Plonky2 (post-quantum) instantiation keeps a constant 157.9\,KB proof and 3.8\,ms verification across depths $L\in\{1,2,4,8,16\}$, at 1.03\,s per delegation step.
Holder-side topology hiding and depth-independent presentation are thus achievable at practical recursive-proving cost.
\end{abstract}

\section{Introduction}
\label{sec:intro}
Transitive authorization is becoming increasingly important in decentralized systems~\cite{augusto2024sok,bartoletti2021sok,defillama}. Modern applications rely on delegation rather than outright transfer of authority: users delegate wallet operations to agents, organizations delegate access rights across services, and DeFi protocols~\cite{werner2022sok,qin2021empirical} delegate financial capabilities while retaining ownership of the underlying assets. Aave's credit delegation illustrates this pattern: a depositor can delegate borrowing capacity to another party without transferring the collateral~\cite{aave}, decoupling the right to act from ownership of the asset. Although current deployments often expose only shallow delegation interfaces, the same pattern naturally extends to settings where capabilities are restricted, re-delegated, and eventually exercised by downstream services across administrative or economic domains. 

Delegatable Anonymous Credentials (DACs)~\cite{chase2006signatures,belenkiy2009randomizable,malleable,crites2019delegatable,camenisch2017practical,mir2023practical,mercurial2024,tdac,SPSEQ-VC} are the natural cryptographic primitive for realizing such transitive authorization.
A DAC lets a holder derive a restricted downstream credential from an existing one, and lets the final presenter prove authorization without revealing unnecessary information about the credential or the delegation path.

\spar{Need for efficient DAC presentations}
In a DAC system, the presentation (or showing) is the use-time operation: after a credential is issued or delegated, every exercise of the delegated capability requires the presenter to convince a verifier that the action is authorized and restriction-preserving~\cite{camenisch2017practical,mir2023practical}.
Unlike one-time issuance or delegation, this cost is incurred on every use.
It may be borne by the holder directly, by wallet software interfacing with decentralized applications~\cite{torres2023your,li2021strong,yu2024don,zhou2023iterative}, or by an outsourced proving service when zero-knowledge generation is expensive~\cite{chiesa2023eos,hu2025dfs,yang2025siniel,abbaszadeh2026single}; in each case it governs the latency, communication, and verification overhead of exercising delegated authority.
Presentation efficiency is therefore not an implementation detail but a precondition for anonymous delegation to be used in practical decentralized applications.
This burden compounds with delegation depth: when a presentation grows with the length of the delegation chain, each additional hop raises the price of every presentation.

Despite the importance of presentation efficiency, no existing DAC achieves a presentation whose cost is independent of the delegation depth.
For early theoretical constructions~\cite{chase2006signatures,belenkiy2009randomizable,malleable}, the credential is an explicit chain, so the presentation proof inherently grows with the delegation depth.
Even for recent schemes~\cite{camenisch2017practical,mir2023practical,mercurial2024,SPSEQ-VC} that are expected to be practical, improving concrete efficiency, strengthening anonymity, or compressing the credential to constant size, the showing remains conditioned on the delegation chain, and unsurprisingly, when the chain is deep, it still carries delegation-dependent material (commitment vectors or witnesses) whose size grows with the depth; relaxing privacy only improves constants.
Moreover, this path-structured material even leaks the delegation topology in addition to its cost, because its size and shape track the delegation path; as such, the holder or an outsourced prover learns the chain depth and per-hop structure even when the verifier-facing presentation is zero-knowledge.
Given that, it becomes essential to make DAC presentations both efficient and topology-hiding, even for deep delegation chains.
\begin{table}[h]
\centering
\caption{Comparison of typical DAC schemes.}
\label{tab:comparison}
\small
\setlength{\tabcolsep}{5pt}
\renewcommand{\arraystretch}{1.25}
\begin{adjustbox}{max width=\columnwidth}
\begin{tabular}{lcccccc}
\toprule
\textbf{Scheme} & \textbf{Ref.} & \textbf{Anon.}\,$^{\dagger}$ & \textbf{Topo.}\,$^{\ddagger}$ & \textbf{PQE}\,$^{\S}$ & $|$\textbf{Pre}$|$\,$^{\P}$ & $|$\textbf{Cred}$|$ \\
\midrule
BCC+09          & \cite{belenkiy2009randomizable} & $\mathcal{S}$   & \xmark & \xmark & $O(L\lambda)$ & $O(L)$ \\
P-UC-DAC        & \cite{camenisch2017practical}   & $\mathcal{W}$   & \xmark & \xmark & $O(L)$         & $O(L)$ \\
Mercurial-DAC   & \cite{crites2019delegatable}    & $\mathcal{W}$   & \xmark & \xmark & $O(L)$         & $O(L)$ \\
TDAC            & \cite{tdac}                     & $\mathcal{W}_L$ & \xmark & \xmark & $O(n)$         & $O(1)$ \\
EQS-DAC         & \cite{mir2023practical}         & $\mathcal{S}$   & \xmark & \xmark & $O(L)$         & $O(1)$ \\
Mercurial-DAC*  & \cite{mercurial2024}            & $\mathcal{S}$   & \xmark & \xmark & $O(L)$         & $O(L)$ \\
FDAAC-CR        & \cite{SPSEQ-VC}                 & $\mathcal{S}$   & \xmark & \xmark & $O(L)$         & $O(1)$ \\
\midrule
\rowcolor{gray!10}
\textbf{PCC (Ours)} & -- & $\mathcal{S}$ & \textbf{\cmark} & \textbf{\cmark} & $\boldsymbol{O(1)}$ & $O(1)$ \\
\bottomrule
\end{tabular}
\end{adjustbox}

\par\vspace{2pt}
\begin{minipage}{0.98\columnwidth}
\footnotesize
\raggedright
\setlength{\parindent}{0pt}
\setlength{\parskip}{1pt}

$^{\dagger}$\,\textbf{Anonymity}:
$\mathcal{S}$ = strong anonymity;
$\mathcal{W}$ = weak anonymity, protecting only verifier-side presentation privacy;
$\mathcal{W}_L$ = level-conditioned weak anonymity, i.e., verifier-side anonymity only among users at the same delegation level $L$.

$^{\ddagger}$\,\textbf{Topo.}: hides delegation topology, including chain depth and per-hop structure, from the holder/prover.

$^{\S}$\,\textbf{PQE}: post-quantum-compatible instantiation under explicit QROM assumptions.

$^{\P}$\,$|$Pre$|$ and $|$Cred$|$ report asymptotic growth mainly with respect to the delegation depth $L$. Attribute/policy-dependent public inputs or witnesses are not counted unless explicitly shown; the full presentation transcript may grow with their encoding. $\lambda$ is the security parameter and $n$ denotes the show-protocol witness size when applicable.

\end{minipage}
\end{table}
\par\smallskip
\spar{\mbox{Why do we need topology hiding?}}
Consider a multi-tier lending service inspired by Aave's credit
delegation~\cite{aaveV3CreditDelegation}. A collateral provider passes
borrowing authority through intermediaries to a manager arranging loans.
The same manager may receive the same rights from the same direct
intermediary via two different upstream paths. Even at equal public
disclosure and permitted local leakage, an explicit credential can reveal
the upstream counterparties and depth to the manager or its prover.
Derecho illustrates how provenance affects transaction
acceptance~\cite{beal2024derecho}: here, surplus lineage combined with
external records could suggest a link to a sensitive jurisdiction and lead
the manager to decline an otherwise eligible loan. Required risk checks and
disclosures remain mandatory. In a tiered fee-sharing arrangement, learning
the depth or identities of upstream intermediaries can also let the manager
seek a shorter channel or bypass a counterparty to retain more fees. The
privacy target is the additional upstream information that the holder's
credential and proving interface would otherwise expose.

\subsection{Practical Obstacles to Efficient DAC Presentations}
Nevertheless, designing a DAC with an efficient presentation is challenging: to prove an action authorized, the protocol must account for every delegation step and every restriction accumulated along the chain, yet doing so without paying a depth-dependent cost or leaking the chain's structure is far from obvious.
Conventional designs settle this by letting the prover carry per-step credential material, trading away exactly the efficiency and topology privacy we seek.

\spar{Practicality issues of earlier theoretical DACs.}
DACs have been studied for nearly two decades, but the early constructions were primarily of theoretical interest due to their path-dependent proof complexity.
The first DAC construction of Chase and Lysyanskaya~\cite{chase2006signatures} established that anonymous credentials could be delegated, but it relied on general non-interactive zero-knowledge proofs for complex NP statements; as later observed by Belenkiy et al.~\cite{belenkiy2009randomizable}, this led to credential-chain representations whose size grows super-polynomially with the delegation depth and hence could not support non-constant-depth delegation in practice.
Belenkiy et al.~\cite{belenkiy2009randomizable} made a substantial theoretical improvement by introducing randomizable proofs and reducing the proof size and proving time for a credential at depth $L$ to $O(L\lambda)$, where $\lambda$ is the security parameter.
However, this still leaves both the proof and the prover's work linear in the delegation depth.
Malleable-signature-based DACs~\cite{malleable} further refined the security definitions and obtained stronger context-hiding guarantees, but their resulting credentials still scale linearly with the number of delegations.
Thus, while these works established the feasibility and formal foundations of strongly private delegation, their authorization evidence remained explicitly path-structured.

This path dependence is the main practicality obstacle in the early theoretical line.
Each delegation contributes proof, signature, or transformation material that must be randomized, transformed, or re-proved when the downstream credential is presented.
Consequently, zero-knowledge hides the chain from the verifier but does not eliminate the chain-shaped material that the prover must carry and process. More recent constructions improve on this foundational line in divergent ways, some relaxing privacy for concrete efficiency, others strengthening privacy with algebraic credentials such as mercurial signatures~\cite{crites2019delegatable,mercurial2024}, threshold credentials~\cite{tdac}, or equivalence-class based schemes~\cite{SPSEQ-VC}, as Table~\ref{tab:comparison} summarizes, none removes the depth dependence from the presentation.

\spar{P-UC-DAC: practicality by exposing lineage.}
P-UC-DAC~\cite{camenisch2017practical} is a major step toward practical DACs: a UC-secure construction with attributes whose pairing-based presentations avoid general-purpose zero-knowledge proofs.
It obtains this efficiency through a presentation-centric privacy target---the presentation hides the credential chain from the verifier, but the delegation interface hands the delegatee the full chain, including the signatures, public keys, and attributes of all previous levels.
The credential thus remains path-structured: a level-$L$ credential is $L$ links, and a presentation proves possession of every link back to the issuer.
P-UC-DAC therefore leaves the presentation depth-dependent and exposes the full delegation lineage to the downstream holder.

\spar{EQS-DAC: strong anonymity, depth-dependent showings.}
More recently, EQS-DAC~\cite{mir2023practical} strengthens privacy by replacing explicit credential chains with SPSEQ-UC signatures over updatable set commitments: each delegation appends one set commitment to a signed commitment vector, and public randomization with key switching makes delegations and showings unlinkable.
This avoids general-purpose zero-knowledge proofs and keeps the credential signature constant-size.
Yet the chain is not removed from the presentation: the showing carries a commitment vector of length $\ell=\Theta(L)$, costing $(\ell+5)|\mathbb{G}_1|+|\mathbb{G}_2|+|\mathbb{Z}_p|$, and the holder additionally stores opening randomness linear in the chain.
Thus EQS-DAC strengthens anonymity over P-UC-DAC, but the showing remains $\Theta(L)$ and the delegation depth stays visible in the holder's own state, so even strong verifier-side anonymity does not hide the topology from the prover.
% More recently, Mir et al.~\cite{mir2023practical} proposed EQS-DAC as a practical DAC construction that strengthens privacy by replacing explicit credential chains with SPSEQ-UC signatures over updatable set commitments.
% At a high level, each delegation appends one set commitment to a signed commitment vector, while public randomization and key switching make delegations and showings unlinkable.
% This design avoids costly general-purpose zero-knowledge proofs and keeps the credential signature itself constant-size.
% However, the chain is not eliminated from the presentation interface; it is encoded as a commitment vector whose length grows with the delegation depth.
% Concretely, for a level-$L$ credential, the showing contains a commitment vector of length growing with $L$, and its communication cost is
% $(k+5)|\mathbb{G}_1|+|\mathbb{G}_2|+|\mathbb{Z}_p|$, where $k$ is the commitment-vector length. Thus, although the credential signature is constant-size, the showing remains $\Theta(L)$ in group elements.
% Moreover, the holder stores commitment-opening randomness linear in the delegation chain.
% Thus, EQS-DAC significantly strengthens privacy over P-UC-DAC, but pays for it with depth-dependent showing bandwidth and holder-side auxiliary state.
% Unfortunately, this still falls short of our target interface, where the presenter should handle a single depth-independent authorization object and the verifier-facing proof should have size and verification time independent of the upstream depth.

\spar{TDAC: closest to depth-independence.}
TDAC~\cite{tdac} comes closest to our target: it compresses the delegated credential into a short decryption key and avoids carrying an explicit delegation-chain witness, so its showing cost is parameterized by the number of attributes rather than the delegation depth.
This independence is not genuine, however. The protocol is level-dependent and anonymity holds only among users at the same level, so the level, i.e.\ the depth, is exposed by construction; and when delegation depth is encoded through accumulated attributes, the showing cost grows with the number of delegated attributes, reintroducing the dependence indirectly.

\spar{Can a DAC presentation be efficient and topology-hiding at once?}
Across these constructions, the obstacle is the same: the prover holds and proves over material whose size and shape mirror the delegation chain, simultaneously inflating the presentation cost and leaking the chain's topology.
Our key observation is that these are not two independent defects but two symptoms of one cause, and can therefore be removed together. Carrying the accumulated authorization in a single depth-independent object, rather than a chain-structured credential, would make both the presentation cost and the exposed topology independent of the delegation depth.

Facing this obstacle in state-of-the-art DACs, we are asking the following question:
\researchquestion{
Can strong anonymity DACs achieve \\
depth-independent presentation and topology-hiding?
}
\subsection{Our Contribution}
We answer this question affirmatively with PCC (Proof-Carrying Credentials), a generic DAC framework whose presentation is depth-independent and topology-hiding while preserving strong anonymity and restricted delegation.
The idea is in the name: rather than carrying the delegation chain, a PCC credential carries a \emph{proof} of it.
Using Non-Uniform Incrementally Verifiable Computation (NIVC), PCC compiles the delegation lineage into a single recursively updatable proof-carrying state that each delegation step extends in place, so a holder retains only its terminal state, local attributes, and a constant-size recursive proof rather than a verifiable list of upstream ancestors or intermediate delegation signatures.
This one representational change removes both limitations at once: a final presentation extends the same state with a presentation predicate, producing a verifier-facing proof whose size and verification cost are independent of the delegation depth, while the holder's local view no longer exposes delegation topology.
Unlike prior constant-size DACs tied to bilinear pairings, PCC is structurally hardness-agnostic and admits an explicit post-quantum-compatible instantiation.
In sum, our contributions are:
\begin{itemize}
\item \textbf{New two-observer privacy notion, Enhanced Delegation Anonymity (EDA), that adds holder-side topology obliviousness to strong anonymity.}
We identify \emph{prover awareness}: across the DAC design space, a holder must process chain-shaped lineage material to present.
Its topology leakage is the privacy face of the same defect that drives depth-dependent presentation cost, so we formalize it as \emph{Enhanced Delegation Anonymity} (EDA): a leakage-parametrized two-observer notion pairing \emph{strong anonymity} over the verifier statement $\mathsf{stmt}_V$ with \emph{holder-side topology obliviousness} over the holder view $\mathsf{View}_H$, the latter relative to an explicit allowed-leakage function $\ell_H(\mathsf{path})$.
Parameterizing on $\ell_H$ casts the informal question of how much structure a holder must learn into the standard simulation/indistinguishability paradigm: we give topology obliviousness as both a simulation definition and an equivalent indistinguishability game over paths with identical allowed leakage.
This is orthogonal to identity anonymity: even BCC+09~\cite{belenkiy2009randomizable}, whose strong anonymity hides delegator identities and unlinkably re-randomizes presentations, fails it, since hiding \emph{who} delegated does not stop the holder from retaining $O(L)$ lineage and learning the depth $L$.
Moreover, the formulation is of independent interest for any delegation or provenance system whose intermediaries should not learn upstream structure, e.g., wallet delegation and service-mediated authorization chains.

\item \textbf{Generic recursive DAC framework, Proof-Carrying Credentials (PCC), that compiles a delegation chain into a single, recursively updatable proof-carrying state, making both presentation cost and exposed topology independent of the delegation depth.}
Following our observation, a PCC holder carries the accumulated authorization as a single proof-carrying state rather than the explicit per-hop history (each level's signature, public key, and attribute set).
We build this over Non-Uniform Incrementally Verifiable Computation (NIVC), whose non-uniformity lets one recursion switch among the heterogeneous issuance, delegation, and presentation predicates.
Each delegation folds one hop into the state by re-verifying the carried proof and checking the new transition, so the holder keeps only its terminal state, local attributes, and an $O(1)$ recursive proof, not the chain.
Validating a depth-$L$ chain thus reduces to verifying one recursive proof, and a final presentation extends the state with a presentation predicate to yield a verifier-facing proof whose size and verification cost are independent of depth.
This single change collapses the $O(L)$ credential and presentation of prior schemes to $O(1)$ and, since the holder no longer holds the chain, delivers EDA's holder-side topology obliviousness as a by-product.

\item \textbf{Co-designed transition predicates and a depth-hiding verification interface that harvests depth-independent, topology-hiding presentation while still preserving restricted delegation, expressive policies, and standard DAC security.}
Collapsing the chain into one opaque proof (Contribution~2) must not sacrifice what makes a DAC useful: enforcement of the accumulated restrictions and support for expressive presentation policies, which verifier-only zero-knowledge cannot deliver\cite{mercurial2024,crites2019delegatable}.
We co-design a predicate family $\Phi=\{\phi_{\mathsf{issue}},\phi_{\mathsf{del}},\phi_{\mathsf{pre}}\}$ enforcing every policy inside the recursive step: $\phi_{\mathsf{del}}$ checks attribute monotonicity $A_{i+1}\subseteq A_i$ and per-hop authorization (restricted delegation), and $\phi_{\mathsf{pre}}$ enforces selective disclosure $A_{\mathsf{pre}}\preceq A_L$ (expressive policies).
Compatibility with topology hiding requires a \emph{depth-hiding} NIVC interface: the verifier checks only the root and terminal states, the recursive proof, and a public terminal condition, leaving the depth and selector sequence proof-internal.
This wrapper adds no hardness assumption beyond the NIVC, only packaging each transition and invoking the NIVC prover over $\Phi$; a wrapper reduction shows it inherits completeness, simulation-extractability, and zero-knowledge, hence correctness, delegation unforgeability, and EDA.
PCC thus harvests depth-independence and topology hiding without surrendering restricted delegation, expressiveness, or security, and remains generic over any depth-hiding NIVC.

\item \textbf{The first DAC to combine a constant-size presentation and credential with a post-quantum-compatible instantiation, implemented over both conventional and post-quantum recursive backends, with an extensive depth-scaling evaluation.}
Because PCC is generic over any depth-hiding NIVC (Contribution~3), we instantiate it in two regimes: a Halo2 backend for the conventional setting and a transparent post-quantum one combining Plonky2-style FRI recursion over Goldilocks, Poseidon hashing, and a hash-based (SPHINCS$^+$-style) signature, analyzed in the QROM without pairings or a trusted setup.
We measure verifier-facing proof size, verification latency, and the per-delegation cost across depths $L\in\{1,2,4,8,16\}$, isolating recursion's role through monolithic-versus-recursive Halo2 and Fractal-versus-Plonky2 comparisons.
The proof size and verification time stay constant across all tested depths ($7.344$\,KB / ${\sim}36$\,ms for Halo2; $157.9$\,KB / $3.8$\,ms for $128$-bit Plonky2); the only depth-dependent work is a per-hop delegation step, paid by the delegator and itself stable across depths (e.g., $1.03$\,s post-quantum), confirming that depth-independence stems from recursion, not the backend.
These results show, for the first time, that holder-side topology hiding and depth-independent presentation are simultaneously achievable at practical recursive-proving cost, including under an explicit post-quantum-compatible instantiation.
\end{itemize}

\section{Models and Technical Overview}
\label{sec:overview}
This section sets up the problem and previews our solution before the formal treatment in Section~\ref{sec:methodology}. We first fix the system and threat model together with the privacy goal (Section~\ref{sub:model}), then state the central design challenge and the techniques PCC uses to meet it (Section~\ref{sub:techniques}), and finally introduce the notation and the single non-standard primitive our construction relies on (Section~\ref{sec:preliminaries}).

\subsection{System and Threat Model}
\label{sub:model}

\begin{figure}[t]
    \centering
    \includegraphics[width=\columnwidth]{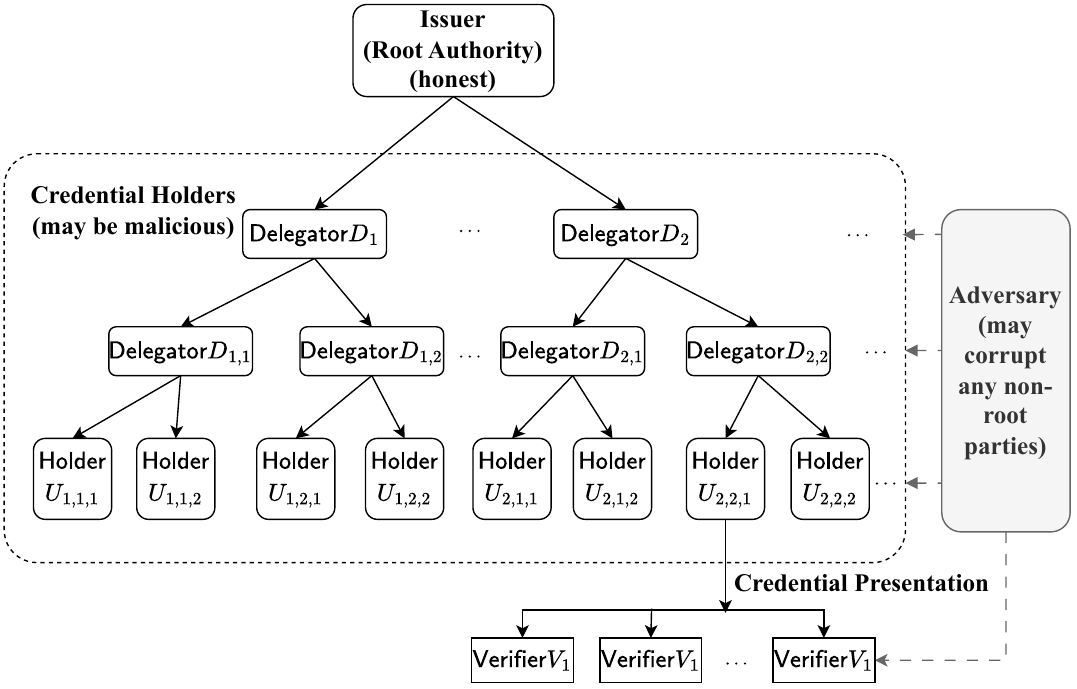}
    \caption{System and threat model. The root authority is honest, while non-root credential holders and verifiers may be malicious and collude. Credential holders may act as delegators, delegatees, or presenters across the issuance, delegation, and presentation phases.}
    \label{fig:system-model}
\end{figure}

\spar{System and roles.}
We consider a hierarchical credential system with a \textbf{Root Authority}, intermediate \textbf{Users}, and external \textbf{Verifiers}; a user acts as \emph{delegator}, \emph{delegatee}, or \emph{presenter} across the \emph{issuance}, \emph{delegation}, and \emph{presentation} phases (Figure~\ref{fig:system-model}). The Root Authority publishes an authenticated root anchor $\RootAnchor$ that verifiers authenticate before accepting presentations; the initial credential state is trusted only once certified by a root issuance signature and verified inside the recursive proof.

\spar{Threat model.}
A PPT adversary may adaptively corrupt users, obtain their keys and credentials, collude, share material, attempt unauthorized delegations, and forge presentations for unauthorized attributes; verifiers may also be malicious. The adversary never obtains $\sk_{\mathsf{root}}$ and cannot make a verifier accept an unauthenticated root anchor, and issuance is answered only under the fixed $\RootAnchor$.

\spar{Privacy goal.}
We target \emph{Enhanced Delegation Anonymity} (EDA), which constrains two observers. A \emph{verifier}, seeing only the presentation transcript, learns nothing beyond the authorized statement and the intentional disclosure (verifier-side anonymity). A \emph{holder}, whose local view comprises the credential, proving interface, and presentation, is not forced to learn delegation-topology metadata beyond an explicit allowed leakage $\ell_H(\mathsf{path})$: beyond $\ell_H$ it cannot cryptographically verify remote ancestors, intermediate credentials or signatures, or the depth $L$ (holder-side topology obliviousness). EDA hides neither application-required disclosures nor the immediate delegation edge, and out-of-band leakage is out of scope. We formalize both notions in Section~\ref{sub:security}.

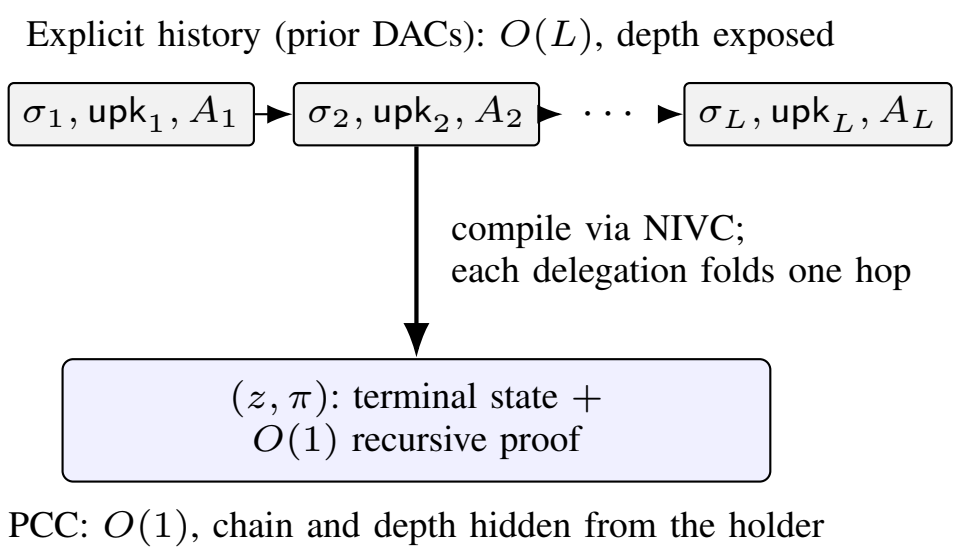
\begin{figure}[ht]
\centering
\begin{tikzpicture}[
  >=Latex, font=\footnotesize,
  link/.style={draw, rounded corners=1pt, fill=black!5, inner sep=2.5pt, font=\scriptsize},
  state/.style={draw, rounded corners=2pt, fill=blue!6, inner sep=4pt, font=\scriptsize, align=center, text width=44mm}
]
\node[link] (l1) {$\sigma_1,\mathsf{upk}_1,A_1$};
\node[link, right=2.5mm of l1] (l2) {$\sigma_2,\mathsf{upk}_2,A_2$};
\node[right=1.5mm of l2] (ld) {$\cdots$};
\node[link, right=1.5mm of ld] (lL) {$\sigma_L,\mathsf{upk}_L,A_L$};
\draw[->] (l1)--(l2); \draw[->] (l2)--(ld); \draw[->] (ld)--(lL);
\node[font=\scriptsize, above=0.5mm of l1.north west, anchor=south west]
  {Explicit history (prior DACs): $O(L)$, depth exposed};

\node[state, below=14mm of l2] (s) {$(z,\pi)$: terminal state $+$ $O(1)$ recursive proof};
\draw[->, thick] (l2.south) -- (s.north)
  node[midway, right=1mm, font=\scriptsize, align=left]
  {compile via NIVC;\\ each delegation folds one hop};
\node[font=\scriptsize, below=0.5mm of s.south]
  {PCC: $O(1)$, chain and depth hidden from the holder};
\end{tikzpicture}
\caption{The key representational change. Prior DACs require the holder to carry and prove the explicit delegation chain, so both its size and the holder's view grow with the depth $L$. PCC compiles the same lineage into a single, recursively updated proof-carrying state $(z,\pi)$: each delegation folds one hop into the state, and the holder retains only the terminal state and a constant-size recursive proof, hiding the chain and the depth.}
\label{fig:pcc-idea}
\end{figure}
\subsection{Challenges and Our Techniques}
\label{sub:techniques}
Realizing EDA with a depth-independent presentation raises three obstacles, which we address in turn.

\spar{Challenge I: a depth-independent, topology-hiding presentation.}
A holder must prove that its authority traces back to the root, yet neither the verifier-facing presentation nor the holder's own view may grow with or reveal the delegation depth. The obvious attempt does not suffice: a zero-knowledge presentation computed over an explicitly stored chain hides the transcript from the verifier, but the holder still stores and proves the per-hop lineage, so its view exposes the depth and the presentation cost grows with $L$. 

\underline{Our approach.} We compile the lineage into a single, recursively updated proof-carrying state (Figure~\ref{fig:pcc-idea}): each delegation folds one hop into the state by re-verifying the carried proof and checking the new transition, so a holder keeps only its terminal state and a constant-size recursive proof, never the chain. We obtain this from incrementally verifiable computation and equip it with a depth-hiding verification interface (Section~\ref{sec:preliminaries}), under which the verifier checks only the root and terminal states, the recursive proof, and a public terminal condition, while the depth and the step sequence stay internal to the proof.

\spar{Challenge II: preserving delegation semantics without the chain.}
Once the explicit lineage is gone, the recursion must still enforce the restrictions accumulated along the path, support expressive presentation policies, and remain unforgeable; moreover it must do so across heterogeneous operations, since issuance, delegation, and presentation enforce different checks. 

\underline{Our approach.} We use a non-uniform variant of incrementally verifiable computation (Definition~\ref{def:nivc}) with a co-designed predicate family $\{\phi_{\mathsf{issue}},\phi_{\mathsf{del}},\phi_{\mathsf{pre}}\}$ that enforces every policy inside the recursive step: the delegation predicate checks attribute narrowing and per-hop authorization (restricted delegation), and the presentation predicate enforces selective disclosure (expressive policies), with the concrete relations given in our construction (Section~\ref{subsec:pcc-construction}). Because the construction merely invokes the recursive prover over this family, it adds no hardness assumption beyond the underlying scheme and inherits its completeness, simulation-extractability, and zero-knowledge, hence correctness, delegation unforgeability, and EDA.

\spar{Challenge III: a uniform and practical recursive step.}
Recursion is the dominant cost, and supporting heterogeneous predicates naively would require three large, independent step circuits. We observe, however, that the three predicates act on a common state $z=\commit(\upk,A;r)$ and are assembled from the same in-circuit gadgets: commitment opening, Poseidon hashing~\cite{poseidon}, hash-based signature verification, and attribute-set relations. They differ in logic, which is precisely why non-uniformity is needed to switch among them, yet they are homogeneous in composition, so the step circuits share a common gadget set and each is dominated by the same Poseidon core. 

\underline{Our approach.} Together with the per-step cost of the recursive scheme, this co-design keeps the marginal cost of one delegation uniform across the three operations and independent of the prior depth, rather than scaling with the chain; we treat this as a property of the predicate family rather than a separate optimization, and report the concrete per-step costs in Section~\ref{sec:eval}.
\subsection{Notation and Primitives}
\label{sec:preliminaries}
We write $[n]=\{1,\ldots,n\}$, use $\negl(\lambda)$ for negligible functions, and $\approx_c$ for computational indistinguishability. We use an EUF-CMA signature scheme $\Sigma$, binding/hiding commitments ($c\leftarrow\commit(m;r)$), and domain-separated hashes $H_{\mathsf{state}},H_{\mathsf{sig}},H_{\mathsf{FS}},H_{\mathsf{pre}}$ with fixed tags. The only non-standard primitive is the following depth-hiding NIVC abstraction.

\spar{Non-Uniform IVC.}
The only non-standard primitive PCC relies on is a depth-hiding non-uniform IVC (NIVC): a recursive proof system $(\setup, \keygen, \prove, \cverify)$ for a family of transition predicates $\F = \{\F_1, \dots, \F_k\}$ over a shared public instance $\rho$, which in PCC contains the root anchor $\RootAnchor$. Each $\prove$ step folds one transition: from a selector $j$, the current and next states $z_i, z_{i+1}$, a witness $w_{i+1}$, and the previous proof $\pi_i$, it outputs a \emph{constant-size} proof $\pi_{i+1}$ attesting that the prior recursion is valid and $\F_j(\rho, z_i, z_{i+1}, w_{i+1}) = 1$; the base step starts from the empty proof $\bot$, and root authenticity is enforced by $\phi_{\mathsf{issue}}$ using the root verification key carried in $\rho$. We assume the standard completeness, simulation-extractability, and zero-knowledge guarantees; the full syntax and these properties appear in Appendix~\ref{app:nivc} (Definition~\ref{def:nivc}).

\spar{Depth-Hiding Verification Interface.}
Some standard presentations of IVC/NIVC expose the step count or selector sequence to the verifier. 
PCC, in contrast, requires a depth-hiding verification interface: the verifier checks only the root state, the terminal state, the recursive proof, and a public terminal condition. 
The actual delegation depth and the non-terminal selector sequence remain hidden inside the recursive proof. 
If a concrete backend requires a fixed recursion horizon,
PCC can be compiled for a public upper bound $L_{\max}$ and padded with dummy transitions; in that case, $L_{\max}$ is public leakage, but the actual delegation depth remains hidden.

%%%%%%%%%%%%%%%%%%%%%%%%%%%%%%%%%%%%%%%%%%%%%%%%%%%%%%%%%%%%%%%%%%%%%%%%%%%%%%%%
\section{Delegatable Anonymous Credentials}
\label{sec:methodology}
In this section, we formally define the syntax and security model for Delegatable Anonymous Credentials (DAC). We model DAC as a hierarchical system where users can issue, delegate, and present credentials. 

\subsection{Syntax of DAC}
Here, $\cred$ denotes an abstract credential object; the syntax does not prescribe whether delegation history is stored explicitly or compressed into a derived proof object. Let $\mathcal{A}$ denote the universe of attributes, and let $\preceq$ denote an application-defined authorization preorder on $\mathcal{A}$; set inclusion is a canonical special case. All verification is relative to an authenticated root anchor $\RootAnchor$. The $\RootAnchor$ binds the root verification key $\ipk_{\mathsf{root}}$ (together with the public parameters); verifiers trust $\RootAnchor$ as the root of authority. A DAC scheme $\Pi_{\mathsf{DAC}}$ consists of the following tuple of polynomial-time algorithms:
\begin{itemize}[noitemsep,leftmargin=*]
    \item $\mathsf{Setup}(1^\lambda) \to \pp$: 
    On input of the security parameter $\lambda$, this algorithm outputs the public parameters $\pp$. We assume $\pp$ is an implicit input to all subsequent algorithms.

    \item $\mathsf{KeyGen}(\pp) \to (\sk, \pk)$: 
    A randomized algorithm that generates a secret key $\sk$ and a corresponding public key $\pk$ for a user or issuer. For the root authority, we write its key pair as $(\isk_{\mathsf{root}}, \ipk_{\mathsf{root}})\leftarrow \mathsf{KeyGen}(pp)$

    \item $\mathsf{Issue}(\isk_{\mathsf{root}}, \RootAnchor, \pk_{\mathsf{user}}, A_{\mathsf{root}}) \to \cred_0$: 
    The root issuance algorithm. It takes as input the root's secret key, the authenticated root anchor, the target user's public key, and an initial attribute set $A_{\mathsf{root}}$. It outputs a level-0 credential $\cred_0$.

    \item $\mathsf{Delegate}(\sk_{\mathsf{parent}}, \cred_{\mathsf{parent}}, \pk_{\mathsf{child}}, A_{\mathsf{child}}) \to \cred_{\mathsf{child}} \cup \{\bot\}$: 
    The delegation algorithm. It takes as input a delegator's secret key, a credential $\cred_{\mathsf{parent}}$ authorizing an attribute set $A_{\mathsf{parent}}$, a delegatee's public key, and a delegated attribute set $A_{\mathsf{child}}$. It enforces monotonicity: if $A_{\mathsf{child}} \preceq A_{\mathsf{parent}}$, it outputs a new credential $\cred_{\mathsf{child}}$; otherwise, it outputs $\bot$.

    \item $\mathsf{Prove}(\sk_{\mathsf{user}}, \cred, m, A_{\mathsf{pre}}) \to \pi$: 
    The presentation algorithm. It takes as input a user's secret key, a valid credential $\cred$, a context message $m$, and a set of attributes $A_{\mathsf{pre}}$ to disclose. It outputs a zero-knowledge proof $\pi$.

    \item $\mathsf{Verify}(\RootAnchor, m, A_{\mathsf{pre}}, \pi) \to \{0, 1\}$: 
    The deterministic verification algorithm. It validates the proof $\pi$ against the authenticated root anchor $\RootAnchor$, the message $m$, and the disclosed attributes $A_{\mathsf{pre}}$. It outputs $1$ for valid proofs and $0$ otherwise. For simplicity, we write the public statement as $(\RootAnchor, m, A_{\mathsf{pre}})$; application-specific context can be folded into $m$ or the disclosed statement without affecting the definitions below.
\end{itemize}

\subsection{Security Definitions}
\label{sub:security}

We separate authorization integrity from the two privacy properties of
Enhanced Delegation Anonymity (EDA). The integrity game adapts the
authorized-lineage and attribute-soundness formulations of
Belenkiy et al.~\cite[Sec.~3.1]{belenkiy2009randomizable},
Bl{\"o}mer and Bobolz~\cite[Defs.~D.1, D.3]{bloemer2018daac}, and
Mir et al.~\cite[App.~D.3]{mir2023practical}.

\begin{definition}[Correctness]
\label{def_correctness}
For any honestly generated, restriction-preserving credential of
authorization $A$ under $\RootAnchor$, a presentation for an eligible
request $(ctx,A_{\mathsf{pre}})$ with $A_{\mathsf{pre}}\preceq A$ verifies
except with negligible probability. In a bounded signature
implementation, the signers must have available signing capacity.
\end{definition}

\begin{definition}[Delegation Unforgeability]
\label{def_unforgeability}
The challenger sets up one authenticated root $\RootAnchor$, gives the
adversary the public parameters, and retains the root signing key. Honest
signing keys in this experiment are dedicated to the instance. The adversary
adaptively accesses the following oracles. A handle $h$ names a recorded
credential, $u$ names a registered holder, and invalid requests return $\bot$.
\begin{itemize}[noitemsep,leftmargin=*]
\item \textbf{New.} Generate an honest holder $u$ and return its public key,
excluding already registered keys and the root key. The adversary can also
register its own public keys subject to the same uniqueness requirement.
\item \textbf{Issue and Delegate.} Issue eligible rights to a recipient, or
delegate from a credential handle of an uncorrupted parent to a recipient,
subject to attribute narrowing. Record the exact signed message and its
transition data, and return a new handle. Give the token to the adversary
if the recipient is adversary-controlled.
\item \textbf{Corrupt.} Release a non-root holder's secret, signer state and
recorded credentials, and classify it as adversary-controlled thereafter.
\item \textbf{Show.} For a valid request to an uncorrupted holder's credential
handle, return a presentation and log the complete canonical verifier
statement $x$ in $\mathcal L_{\mathsf{show}}$.
\end{itemize}
Let $\mathsf{AuthCl}_{\mathcal A}$ be the downward closure under $\preceq$
of rights in credentials delivered to adversary-controlled recipients or
exposed through corruption. The adversary wins if it outputs an accepting
presentation under the fixed root for disclosure $A^*$ with
$x^*\notin\mathcal L_{\mathsf{show}}$ and $A^*\notin\mathsf{AuthCl}_{\mathcal A}$.
Freshness excludes replays with different proof bytes. The verifier matches
the public statement to the expected request (which can include a fresh
nonce). The winning probability must be negligible.
\end{definition}

To formalize privacy notations, let \(\rho=\RootAnchor\) be fixed and
let \(ctx\) encode the public presentation context, including any policy.
The challenger gives the adversary honestly generated public parameters.
In each game it samples a uniform bit \(b\) after the adversary has
chosen the candidates and obtained its auxiliary view \(u\), which
is shared between both worlds. Security means negligible advantage
\(\left|\Pr[b'=b]-1/2\right|\) for each PPT adversary's guess \(b'\).

\begin{definition}[Verifier-Side Anonymity]\label{def_verif_Anony}
The adversary supplies two valid paths under \(\rho\) and their
presentation inputs (or references to challenger-held inputs), both
authorizing the same public statement \(\mathsf{stmt}_V\).
In PCC this is \((\rho,z_{\mathsf{root}},z_{\mathsf{pre}},\tau_{\mathsf{pre}})\),
determined by \((\rho,ctx,A_{\mathsf{pre}})\).
Any intentional application disclosures must also match;
PCC itself outputs neither a holder key nor a pseudonym.
The challenger honestly produces a fresh presentation from path \(b\)
and gives its full serialized public transcript to the adversary,
who guesses \(b\). The auxiliary view \(u\) can include candidate
histories, credentials, and authorization keys, but not the challenge
prover's fresh coins.
\end{definition}
This game adapts left--right DAC anonymity (e.g., Mir et
al.~\cite[App.~D.3, Fig.~6]{mir2023practical}) to standalone presentation;
it does not model anonymity of the delegation interaction.
Application disclosures can themselves identify or link presentations.

\begin{definition}[Holder-Side Topology Obliviousness]\label{def_holder_Anony}
The adversary supplies two valid upstream histories, possibly
adversarially prepared, each extendable by an honest final handoff
to the same holder \(H\). The candidates must agree on the allowed leakage
\[
\ell_H=(\rho,\ell_{\mathsf{loc}},\ell_{\mathsf{pre}},
          \ell_{\mathsf{edge}},\ell_{\mathsf{sys}}).
\]
Here \(\ell_{\mathsf{loc}}\) contains \(H\)'s own keys and terminal
attributes; \(\ell_{\mathsf{pre}}\), common presentation inputs;
\(\ell_{\mathsf{edge}}\), exposed immediate-edge information; and
\(\ell_{\mathsf{sys}}\), public parameters and any fixed public depth bound.
It does not directly include the actual depth or remote lineage.
The challenger runs the final handoff for branch \(b\) with fresh
randomness and honestly runs the presentation. The adversary, observing as
\(H\), receives the full view \(\mathsf{View}_H\): the credential, local
proving inputs, visible coins and API metadata, and the presentation.
It then guesses \(b\).
Its auxiliary view \(u\) can include both upstream histories and their
secrets, but not the handoff prover's private coins for this challenge
or messages revealing the selection outside the returned view. Physical
side channels are out of scope.
\end{definition}
In PCC the view includes \(\pcred_H=(\rho,\token_H)\) with
\(\token_H=(z_H,\pi_H,\upk_H,A_H,r_H)\), and uses the presentation
witness \(w_{\mathsf{pre}}=(\usk_H,ctx,A_{\mathsf{pre}},A_H,r_H)\).
The game tests the full view, including proof serialization and parser
metadata. A simulator of the joint view from the common leakage and
auxiliary view suffices for this game.

\spar{Enhanced Delegation Anonymity}
A DAC satisfies EDA if it meets both definitions. Hiding identities
does not imply hiding topology: Belenkiy et
al.~\cite[Sec.~1]{belenkiy2009randomizable} hide upstream identities but
carry a chain whose size grows with depth. Our holder-side guarantee
does not hide disclosures in \(\ell_H\) or prevent an informed sender
from deliberately revealing the selected route.

\section{Proof-Carrying Credentials}
\label{sec:construction}
We introduce \emph{Proof-Carrying Credentials} (PCC), a recursive DAC structure that compiles delegation validity into a proof-carrying state. Section~\ref{subsec:existing-solutions} reviews a representative signature-and-NIZK baseline whose explicit history makes both the validation data and the exposed lineage grow with the delegation depth; Section~\ref{subsec:pcc-construction} replaces this history with a single recursively updated state built over Non-Uniform IVC (NIVC)~\cite{valiant2008ivc}, giving a constant-size verifier-facing object for paths of arbitrary depth~\cite{chiesa2024security} that hides remote delegation topology. The PCC wrapper inherits the completeness, simulation-extractability, and zero-knowledge of the underlying NIVC, yielding correctness, delegation unforgeability, and Enhanced Delegation Anonymity at the DAC layer.
\subsection{A Signature-and-NIZK Baseline}
\label{subsec:existing-solutions}
We first recall the explicit-history pattern underlying representative
signature-and-NIZK DAC constructions, including P-UC-DAC~\cite{camenisch2017practical}.
In such a construction, a level-$L$ holder receives a credential containing
the entire delegation path:
\begin{equation}
    \credl =
    \left(
        \{(\sigma_i,\upk_i,A_i)\}_{i=1}^{L},
        \usk_L
    \right),
    \label{equ:credential-L}
\end{equation}
where each $\sigma_i$ authorizes the transition from $\upk_{i-1}$ to
$\upk_i$ and binds the delegated attribute set $A_i$.

To present the credential, the holder proves in zero knowledge that the
stored path is valid. Setting $\upk_0=\ipk_{\mathsf{root}}$, the presentation relation
contains the following depth-dependent checks:
\begin{equation*}
\label{eq:baseline-relation}
\begin{gathered}
\Gamma_i :=
\left[
\begin{aligned}
&\Sigma.\cverify(\upk_{i-1},(\upk_i,A_i),\sigma_i)=1\\
&{}\land A_i \preceq A_{i-1}
\end{aligned}
\right],\\
\relation_{\mathsf{hist}}
=
\left\{
\begin{aligned}
(x,w):\ &\bigwedge_{i=1}^{L}\Gamma_i 
\land \upk_L=\PK(\usk_L)\\
&{}\land A_{\mathsf{pre}}\preceq A_L
\end{aligned}
\right\}.
\end{gathered}
\end{equation*}
Here the public statement includes the issuer key and the disclosed
presentation statement, while the witness contains the path, the holder's
secret key, and the attribute sequence.

This construction hides the delegation path from the verifier through zero knowledge, but the holder must still store and process the full chain $\{(\sigma_i,\upk_i,A_i)\}_{i=1}^{L}$ to present, so its local view stays explicitly history-dependent and retains cryptographically verifiable lineage. PCC eliminates this explicit history by replacing the path with a recursively updated proof-carrying state.

\subsection{PCC Construction}
\label{subsec:pcc-construction}
\begin{figure*}[t]
    \centering
    \includegraphics[width=\textwidth]{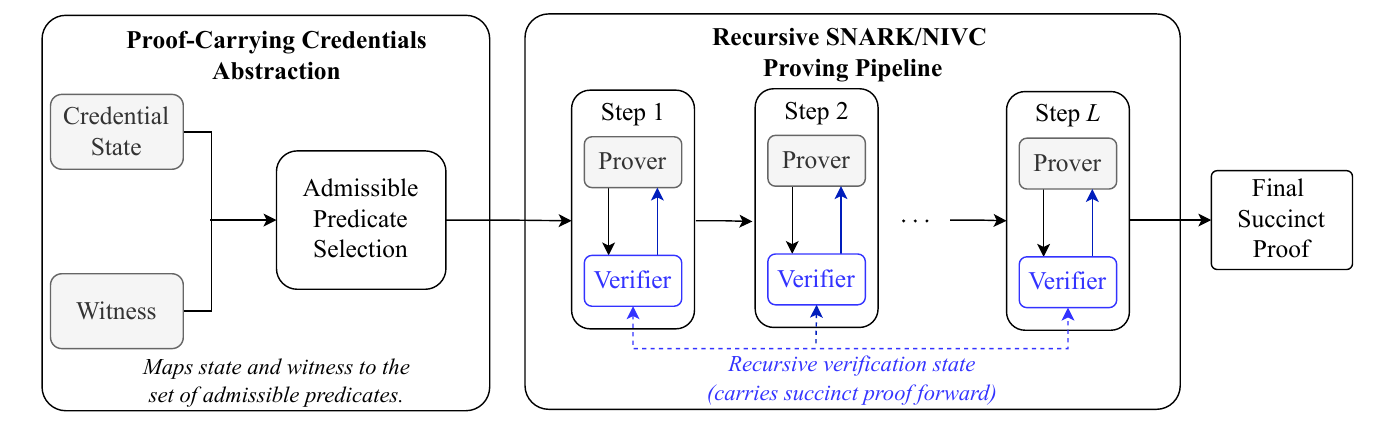}
    \caption{PCC workflow based on NIVC. }
    \label{fig:NIVC}
\end{figure*}
Our proof-carrying credential construction leverages Non-Uniform Incrementally Verifiable Computation (NIVC), recalled in Definition~\ref{def:nivc}, to realize a non-interactive recursive delegation model. 
The non-uniform feature is necessary because the recursive computation must support heterogeneous transition types: whereas standard IVC repeatedly applies a single transition function, our construction must switch among issuance, delegation, and presentation predicates.
By running NIVC, the validity of the delegation path is accumulated into a recursively updated proof state carried by the credential.
This enables a holder to present the credential together with a proof that the delegated authorization is valid with respect to a lineage rooted at the issuer.

We define three state-transition predicates,
$\Phi=\{\phi_{\mathsf{issue}},\phi_{\mathsf{del}},\phi_{\mathsf{pre}}\}$, for PCC. These predicates encode the core security policies and validity constraints enforced by the respective step circuits at each transition.
Before defining the predicates, we formalize the state and witness structures.
\begin{itemize}[noitemsep,leftmargin=*]
    \item \textbf{State ($z_i$):} The state at step $i$ is a commitment $z_i=\commit(\upk_i,A_i;r_i)$ to the current delegatee's verification key $\upk_i$ and authorized attribute set $A_i$. The public initial state is derived from the authenticated anchor as
$z_{\mathsf{root}}:=H_{\mathsf{state}}(\mathsf{tag}_{\mathsf{issue}},\RootAnchor)$.
    \item \textbf{Witness ($w_{i+1}$):}
Write $o_i=(\upk_i,A_i,r_i)$ for the opening of $z_i$.
The witness contains the private inputs for the transition:
\[
\small
        w =
    \begin{cases}
    (\sigma_{\mathsf{root}},o_0,\mathsf{meta}_{\mathsf{iss}})
        & \text{ issue},\\
    (\sigma_i,o_{i+1},o_i,\mathsf{meta}_{\mathsf{del}})
        & \text{ del},\\
    (\usk_i,ctx,A_{\mathsf{pre}},A_i,r_i)
        & \text{ pre}.
    \end{cases}
\]
\end{itemize}
\spar{The Issuance Predicate}
The predicate $\phi_{\mathsf{issue}}$ initializes the recursive lineage.
The issuer verification key is resolved from the authenticated anchor
and bound to the fixed instance $\rho$.
It checks that $z_{\mathsf{root}}=H_{\mathsf{state}}(\mathsf{tag}_{\mathsf{issue}},\RootAnchor)$,
that $z_0=\commit(\upk_0,A_0;r_0)$, and that
\[
\signverify(\ipk_{\mathsf{root}},
H_{\mathsf{sig}}(\mathsf{tag}_{\mathsf{issue}},z_0,A_0,\mathsf{meta}_{\mathsf{iss}}),
\sigma_{\mathsf{root}})=1 .
\]
\spar{The Delegation Predicate} 
The predicate $\phi_{\mathsf{del}}$ governs the extension of the delegation path (function selector $j=\mathsf{del}$). It takes as public inputs the current state commitment $z_{i}$ and the next state commitment $z_{i+1}$, along with the witness $w_{i+1}$. The step circuit accepts the transition if and only if $\phi_{\mathsf{del}}(\rho,z_i,z_{i+1},w_{i+1}) = 1$. The constraints are defined as follows:
\begin{itemize}[noitemsep,leftmargin=*]
    \item \textit{Consistency:} The witness $w_{i+1}$ must correctly open the commitments $z_{i}$ and $z_{i+1}$ to reveal the underlying states $(\upk_i, A_i)$ and $(\upk_{i+1}, A_{i+1})$.
    \item \textit{Attribute Hierarchy Policy:} It enforces the monotonicity of authority by verifying that the delegated attributes are a valid subset of the delegator's attributes:
        $
            A_{i+1} \subseteq A_{i}.
        $
    \item \textit{Transition Authorization:} It verifies that the transition from $z_i$ to $z_{i+1}$ is cryptographically authorized by the owner of the current state. This is enforced by checking that $\sigma_i$ (included in $w_{i+1}$) is a valid signature on the domain-separated delegation message $M_i^{\mathsf{del}}$ under $\upk_i$:
       \begin{equation}
            \begin{aligned}
            M_i^{\mathsf{del}}
            &:= H_{\mathsf{sig}}\big(
                \mathsf{tag}_{\mathsf{del}}, z_i, z_{i+1}, A_{i+1}, \mathsf{meta}_{\mathsf{del}}
            \big), \\
            &\signverify(\upk_i,M_i^{\mathsf{del}},\sigma_i)
            =1 .
            \end{aligned}
        \end{equation}
\end{itemize}

\spar{The Presentation Predicate} For the terminal presentation step, we write
$i=L$ and use $(\upk_L,A_L,\usk_L)$ to denote the terminal holder state.
The predicate $\phi_{\mathsf{pre}}$, corresponding to selector
$j=\mathsf{pre}$, finalizes a credential into a context-bound presentation.
It takes the terminal credential-state commitment $z_L$ as the current
state and a public presentation state $z_{\mathsf{pre}}$ as the next state.
The function outputs $1$ if and only if:

\begin{itemize}[noitemsep,leftmargin=*]
    \item \textbf{Credential-state consistency.}
    The witness opens the terminal credential state $z_L$ to
    $(\upk_L,A_L,r_L)$.

    \item \textbf{Presentation binding.}
    The output state is the deterministic public binding
    $
        z_{\mathsf{pre}}
        =
        H_{\mathsf{pre}}(\mathsf{pre},ctx,A_{\mathsf{pre}}),
    $
    where $ctx$ is the presentation context and $A_{\mathsf{pre}}$ is the
    disclosed attribute set.

    \item \textbf{Ownership proof.}
    The circuit checks $\mathsf{KeyRel}(\upk_L,\usk_L)=1$
    (Definition~\ref{def:pcc-key-relation}).

    \item \textbf{Selective disclosure logic.}
    The disclosed attributes are authorized by the terminal credential:
    $
        A_{\mathsf{pre}} \preceq A_L .
    $
\end{itemize}
Based on the NIVC primitive and the predicates defined above, we now formally construct the Proof-Carrying Credentials scheme $\pcc$.
\begin{remark}
  The integrity of the historical path leading to $z_{i}$ (lineage validation) is guaranteed by the underlying NIVC recursive verification mechanism and is implicitly checked before executing this step.  
\end{remark} 

\begin{definition}[Proof-Carrying Credentials Scheme]
\label{def:pcc_scheme}
The Proof-Carrying Credentials scheme is a tuple of polynomial-time algorithms
\[
\begin{aligned}
\pcc = (&\pccsetup,\pcckeygen,\\
        &\pccderive,\pccverify)
\end{aligned}
\]
instantiated over the NIVC primitive (Definition~\ref{def:nivc}) and the predicate set $\Phi=\{\phi_{\mathsf{issue}},\phi_{\mathsf{del}},\phi_{\mathsf{pre}}\}$:
\begin{itemize}[noitemsep,leftmargin=*, align=left]
    \item $\pccsetup(1^\lambda) \rightarrow \pp$: 
    The generator runs the NIVC setup $\nivcsetup(1^\lambda)$ to produce public parameters $\pp$.
    \item $\pcckeygen(\pp) \rightarrow (\proverk, \verifierk)$: 
    The indexer takes the parameters and the specific logic description $\Phi$ to generate the proving and verification keys:
    $
        (\proverk, \verifierk) \leftarrow \nivckeygen(\pp, \Phi)
    $
    
    \item \textbf{Auxiliary subroutine} $\pccderive^*$:
    
    Given$(\proverk, j, \rho, z_{\mathsf{root}}, z_i, z_{i+1}, w_{i+1}, \pi_i),$
    this internal subroutine updates the recursive proof from state $z_i$
    to state $z_{i+1}$ under the selected predicate
    $j\in\{\mathsf{issue},\mathsf{del},\mathsf{pre}\}$. It invokes the underlying NIVC
    prover as
    \[
        \pi_{i+1}
        \leftarrow
        \nivcprove(
            \proverk,
            j,
            \rho,
            z_{\mathsf{root}},
            z_i,
            z_{i+1},
            w_{i+1},
            \pi_i
        ).
    \]
    It returns the updated recursive proof $\pi_{i+1}$.

    In delegation mode, $j=\mathsf{del}$ and $z_{i+1}$ is the next
    credential-state commitment. In presentation mode, $j=\mathsf{pre}$
    and $z_{i+1}=z_{\mathsf{pre}}$ is the public presentation-binding
    state.
   \item \textbf{Wrapper Syntax.} To streamline the DAC lifecycle, we introduce the high-level wrapper $\pccderive$, which encapsulates witness construction and automates predicate selection. Here and below, $\token_i$ denotes the credential token carried at step $i$.
    \item  $\pccderive(\rho,z_{\mathsf{root}}, \token_i, \mathsf{args}; \sk)\rightarrow \token':$

    The wrapper selects the transition predicate according to $\mathsf{args}$.
    In issuance mode $\token_i=\bot$; otherwise it parses
    $\token_i=(z_i,\pi_i,\upk_i,A_i,r_i)$.
    \begin{itemize}[noitemsep,leftmargin=*]
        \item \textbf{Issuance mode.}
            If $\mathsf{args}$ encode $(\upk_0,A_0,\mathsf{meta}_{\mathsf{iss}})$,
            sample $r_0$, set $z_0:=\commit(\upk_0,A_0;r_0)$, and compute
            \[
            \begin{aligned}
            M_{\mathsf{iss}}
            &:= H_{\mathsf{sig}}(
                \mathsf{tag}_{\mathsf{issue}}, z_0, A_0,
                \mathsf{meta}_{\mathsf{iss}}),\\
            \sigma_{\mathsf{root}}
            &\leftarrow \Sigma.\mathsf{Sign}(\isk_{\mathsf{root}},M_{\mathsf{iss}}),\\
            w_0
            &:= (\sigma_{\mathsf{root}},\upk_0,A_0,r_0,\mathsf{meta}_{\mathsf{iss}}).
            \end{aligned}
            \]
            Then compute
            \[
            \begin{aligned}
            \pi_0 &\leftarrow \pccderive^*(\proverk,\mathsf{issue},\rho,\\
            &\qquad z_{\mathsf{root}},z_{\mathsf{root}},z_0,w_0,\bot),
            \end{aligned}
            \]
            and output
            \[
            \token_0:=(z_0,\pi_0,\upk_0,A_0,r_0).
            \]
        \item \textbf{Delegation mode.}
        If $\mathsf{args}$ encode delegation parameters
        $(\upk_{i+1},A_{i+1},\mathsf{meta}_{\mathsf{del}})$, set
        $j=\mathsf{del}$, sample $r_{i+1}$, and compute
        $
            z_{i+1}:=\commit(\upk_{i+1},A_{i+1};r_{i+1}).
        $
        Define the delegation message and signature as
        \[
        \begin{aligned}
        M_i^{\mathsf{del}}
        &:= H_{\mathsf{sig}}(\mathsf{tag}_{\mathsf{del}},z_i,z_{i+1},
        A_{i+1},\mathsf{meta}_{\mathsf{del}}),\\
        \sigma_i
        &\leftarrow \Sigma.\mathsf{Sign}(\usk_i,M_i^{\mathsf{del}}).
        \end{aligned}
        \]
        Construct the delegation witness
        \[
            w_{i+1}:=(\sigma_i,o_{i+1},o_i,\mathsf{meta}_{\mathsf{del}}),
        \]
        and compute
        \[
            \begin{aligned}
            \pi_{i+1} &\leftarrow \pccderive^*(\proverk,\mathsf{del},\rho,\\
            &\qquad z_{\mathsf{root}},z_i,z_{i+1},w_{i+1},\pi_i).
            \end{aligned}
        \]
        Output$\token_{i+1}:=(z_{i+1},\pi_{i+1},\upk_{i+1},A_{i+1},r_{i+1}). $

        \item \textbf{Presentation mode.}
        If $\mathsf{args}$ encode presentation parameters
        $(ctx,A_{\mathsf{pre}})$, set $j=\mathsf{pre}$ and compute
        \[
            z_{\mathsf{pre}}
            :=
            H_{\mathsf{pre}}(\mathsf{pre},ctx,A_{\mathsf{pre}}).
        \]
        Construct the presentation witness
        \[
            w_{i+1}:=
            (\usk_i,ctx,A_{\mathsf{pre}},A_i,r_i),
        \]
        and compute
        \[
            \begin{aligned}
            \pi_{i+1} &\leftarrow \pccderive^*(\proverk,\mathsf{pre},\rho,\\
            &\qquad z_{\mathsf{root}},z_i,z_{\mathsf{pre}},w_{i+1},\pi_i).
            \end{aligned}
        \]
        Output the final presentation token
        \[
            \mathsf{Token}_{\mathsf{pre}}
            :=
            (z_{\mathsf{pre}},\pi_{i+1},ctx,A_{\mathsf{pre}}).
        \]
    \end{itemize}
    \item \textbf{Verification.}
Given $(\verifierk,\rho,\token_{\mathsf{pre}})$, parse
\[
    \token_{\mathsf{pre}}
    = (z_{\mathsf{pre}},\pi_{\mathsf{pre}},ctx,A_{\mathsf{pre}}).
\]
Check $ctx$ and $A_{\mathsf{pre}}$ against the expected request and set
$z_{\mathsf{root}}:=H_{\mathsf{state}}(\mathsf{tag}_{\mathsf{issue}},\rho)$.
Recompute the expected presentation state
$
    \widehat z_{\mathsf{pre}}
    :=
    H_{\mathsf{pre}}(\mathsf{pre},ctx,A_{\mathsf{pre}})
$
and rejects if $\widehat z_{\mathsf{pre}}\neq z_{\mathsf{pre}}$.
It then forms the public terminal condition
$
    \tau_{\mathsf{pub}}
    :=
    (\mathsf{tag}=\mathsf{pre},ctx,A_{\mathsf{pre}})
$
and checks
\[
    b \leftarrow
    \nivcverify
(\verifierk,\rho,z_{\mathsf{root}},z_{\mathsf{pre}},
\pi_{\mathsf{pre}},\tau_{\mathsf{pub}}).
\]
The verifier accepts iff $b=1$. The verifier is not given the delegation
depth, intermediate credentials, intermediate states, or the non-terminal
selector sequence.
\end{itemize}
\end{definition}
We require $\pcc$ to satisfy completeness, simulation extractability, and zero-knowledge.
For delegation steps, we write the concrete credential at step $i$ as $\pcred_i := (\rho,\token_i),$ where  $\token_i=(z_i,\pi_i,\upk_i,A_i,r_i).$
For presentation, the final token exposes the public projection $(z_{\mathsf{pre}},\pi)$ together with the presentation context encoded in $\mathsf{args}$.
\subsection{Security Analysis}
\label{subsec:pcc-security}

The representation by itself does not imply privacy or integrity. The
recursive verifier must enforce the execution language
$\mathsf{issue}\,\mathsf{del}^{*}$ for a carried credential and
$\mathsf{issue}\,\mathsf{del}^{*}\,\mathsf{pre}$ for a showing. In
particular, the base proof is accepted only by the issuance branch, every
successor extends an accepting credential under the same root, and a
presentation cannot be used as a predecessor. Appendix~\ref{app:nivc}
specifies the proof interface. We assume canonical input encodings, binding
state commitments, and collision resistance for the signed and presentation
transcripts.

\begin{theorem}[Conditional Security of PCC]
\label{thm:pcc-security}
Let the backend satisfy completeness, adaptive auxiliary-input zero
knowledge and fresh-statement trace extraction after simulated showings
(Definitions~\ref{def:nivc} and~\ref{def:pcc-proof-interface}). Let the
authorization signature be EUF-CMA secure, and let the ownership
relation satisfy Definition~\ref{def:pcc-key-relation}. Under the binding
conditions above, the induced DAC is correct and delegation-unforgeable.
Auxiliary-input zero knowledge of the presentation proof yields
verifier-side anonymity for a fixed public statement. With
credential-proof zero knowledge and a canonical holder representation
apart from the credential proof bytes determined by the allowed leakage,
PCC also satisfies holder-side topology obliviousness at an honest final
handoff.
\end{theorem}

\begin{proof}[Proof overview]
Completeness follows by induction over authorized transitions. In the
integrity game, replace honest showings by simulations, then extract from
a fresh accepting output. Match the trace from the root against the signing
ledger. A first unmatched transition under an honest signer yields a
signature forgery or a binding violation. Reaching an adversarial holder
through recorded edges puts the final disclosure in the authorized
closure. Otherwise the extracted terminal secret for a recorded honest
holder enables a fresh signature under that holder's key. For privacy, the
verifier proof is simulated from its public statement. The holder proof is
simulated from its fixed credential statement; the showing is then generated
using that proof and the available terminal witness. The reductions appear
in Appendix~\ref{app:pqdac-unforge} and Appendix~\ref{app:pqdac-eda}.
\end{proof}
\section{Post-Quantum Instantiation of PCC}
\label{sec:inst}
In this section, we present a concrete post-quantum instantiation of PCC.
The generic PCC framework is structurally hardness-agnostic: it yields post-quantum security provided that the underlying signature scheme and recursive proof system remain secure against quantum adversaries.
Our goal here is therefore not to redefine PCC, but to instantiate it with primitives that simultaneously support post-quantum security and efficient recursive proving.

To maximize circuit efficiency in the concrete realization, we prioritize primitives that minimize arithmetic complexity inside the recursive proof system.
Our primary instantiation combines Plonky2-style FRI-based recursion, Poseidon hashing, and a circuit-oriented hash-based authorization scheme.
This preserves the generic DAC interface of Section~\ref{sec:construction}.
The 100- and 128-bit settings reported in the original evaluation are
nominal parameter targets, not security levels established by this analysis.

\subsection{Cryptographic Primitives and Parameters}
\label{subsec:primitives}
Our post-quantum-compatible instantiation combines transparent recursive
proving with circuit-friendly authorization checks: Goldilocks arithmetic
($p=2^{64}-2^{32}+1$, FRI over $\Fp^{2}$) enables efficient Plonky2-style FRI
recursion, Poseidon provides arithmetization-friendly hashing, and a
Poseidon-based, SPHINCS\textsuperscript{+}-style~\cite{SPHINCS} hash-based signature
authorizes issuance and delegation without pairing- or discrete-log
assumptions. We retain Fractal-style transparent recursion as a comparison
point in Section~\ref{sec:eval}. The full primitive stack, parameters, and
security rationale (Table~\ref{tab:pq-primitives}) are deferred to
Appendix~\ref{app:pq-primitives}, under whose explicit QROM assumptions the
instantiation is post-quantum-compatible.

\subsection{Protocol Design of PQDAC}
\label{sec:protocol}

We instantiate a concrete post-quantum DAC, $\Pi_{\mathsf{PQDAC}}$, as the generic
$\Pi_{\mathsf{PCC}}$ of Definition~\ref{def:pcc_scheme} run over the primitive
stack of Table~\ref{tab:pq-primitives}: its algorithms are exactly
$\pccsetup/\pcckeygen/\pccderive/\pccverify$ with $\Sigma$ and the NIVC
instantiated post-quantumly, and user keys are signature keypairs
($\usk:=\sk_{\mathsf{sig}}$, $\upk:=\pk_{\mathsf{sig}}$). The full algorithmic
specification is given in Figure~\ref{fig:pqdac_protocol}. The protocol is parameterized by
$\Phi=\{\phi_{\mathsf{issue}},\phi_{\mathsf{del}},\phi_{\mathsf{pre}}\}$, where
$\phi_{\mathsf{issue}}$ authenticates root issuance, $\phi_{\mathsf{del}}$
governs recursive delegation, and $\phi_{\mathsf{pre}}$ governs context-bound
presentation. A credential evolves through three phases:
\begin{itemize}[noitemsep,leftmargin=*]
    \item \textbf{Setup and issuance.} $\textsf{Setup}$ compiles $\Phi$ into
    $(\proverk,\verifierk)$; the root authority invokes $\pccderive$ in issuance
    mode under $\RootAnchor$, producing $\pcred_0=(\rho,\token_0)$ with
    $\rho:=\RootAnchor$.
    \item \textbf{Recursive delegation.} $\textsf{Delegate}$ invokes $\pccderive$
    in delegation mode on
    $(\pcred_i,\upk_{i+1},A_{i+1},\mathsf{meta}_{\mathsf{del}},\usk_i)$, yielding
    $\token_{i+1}$ with a constant-size verifier-facing proof component.
    \item \textbf{Presentation.} $\textsf{Present}$ invokes $\pccderive$ in
    presentation mode for $\mathsf{ctx}:=(m,\policy)$ and $A_{\mathsf{pre}}$;
    $\textsf{Verify}$ checks $\token_{\mathsf{pre}}$ against $\RootAnchor$, from
    which $z_{\mathsf{root}}$ is derived.
\end{itemize}

The realization inherits security only if it implements the interfaces in
Theorem~\ref{thm:pcc-security}. In particular, security of isolated
FRI/Fiat--Shamir proofs does not establish extraction of a full recursive
trace after simulated showings.

\begin{theorem}[Conditional Delegation Unforgeability]
\label{thm:pqdac-unforge}
If the realization satisfies the integrity premises of
Theorem~\ref{thm:pcc-security}, it is delegation-unforgeable
(Definition~\ref{def_unforgeability}).
\end{theorem}

\begin{theorem}[Conditional Enhanced Delegation Anonymity]
\label{thm:pqdac-eda}
If the realization satisfies the privacy premises of
Theorem~\ref{thm:pcc-security}, it satisfies EDA.
\end{theorem}

\begin{theorem}[Conditional Post-Quantum Security]
\label{thm:pqdac-qrom}
The same conclusions hold against quantum polynomial-time adversaries
with classical protocol queries when all interfaces of
Theorem~\ref{thm:pcc-security} hold in the same quantum security model.
Appendix~\ref{app:qrom-model} specifies the assumptions and their scope.
\end{theorem}

\section{Evaluation}
\label{sec:eval}
\subsection{Experimental Setup}
\label{sec:experiment_setup}
\spar{Environment and metrics.}
All benchmarks were executed on a desktop workstation with an AMD Ryzen 7
9800X3D processor, 32\,GB of RAM, and a 1\,TB SSD, running Ubuntu 22.04 LTS.
Unless otherwise stated, measurements were collected in single-threaded mode
over 10 independent trials, and we report the mean. We evaluate three
metrics: final verifier-facing proof size, final verification latency, and
delegation time per recursive step. For recursive DACs, the per-step
delegation time is the marginal online cost of extending an existing
credential by one hop, rather than the cumulative cost of reconstructing the
full delegation chain. All proof sizes reported in this section refer to the
final object delivered to the verifier.

\spar{Workloads.}
We evaluate delegation depths up to $L=16$, which is conservative relative
to typical transitive-authorization deployments: DeFi credit delegation often
uses small hop counts, while SSI, enterprise IAM, and service-mediated
authorization chains rarely exceed single-digit or low double-digit depth.
For the Halo2 experiments, we use a workload with 20 attributes and evaluate
both selective disclosure and richer validity predicates. For the
post-quantum experiments, we report results at 100-bit and 128-bit security.
Unless otherwise stated, compared systems use matched delegation depth,
attribute-set size, and presentation-policy complexity.

\spar{Baselines.}
For the non-post-quantum evaluation, we compare against
\textbf{P-UC-DAC}~\cite{camenisch2017practical}, the closest prior practical
proof-based DAC baseline. To isolate the role of recursion, we also compare \textbf{Halo2 Monolithic} with \textbf{Halo2 Recursive} under the same backend and matched workload: the former encodes accumulated lineage-validation checks directly in the final presentation circuit, whereas the latter carries lineage validity recursively through the PCC state and exports a single verifier-facing proof object. For the post-quantum
setting, we compare a non-recursive Fractal-style baseline, a recursive
Fractal baseline, and our Plonky2-based recursive PCC. We do not use EQS-DAC
as a primary experimental baseline because it belongs to a different design
line that avoids general proof-system machinery; we retain it as a
qualitative comparison point in Table~\ref{tab:comparison}.

\spar{Evaluation goals.}
This section answers four questions:
\begin{itemize}[leftmargin=*,noitemsep]
    \item \textbf{RQ1:} Compared with the closest prior practical proof-based DAC baseline, does PCC remove depth-dependent verifier-facing presentation growth while incurring only a moderate constant recursive overhead?
    \item \textbf{RQ2:} On the Halo2 backend, does the PCC turn recursive lineage carrying into a constant-size verifier-facing proof interface?
    \item \textbf{RQ3:} Does the post-quantum instantiation remain practical
    at deployment-realistic delegation depths?
    \item \textbf{RQ4:} Under matched FRI configurations, how do PCC's
    presentation latency and peak resident memory vary with delegation
    depth compared with exact-depth and fixed-bound explicit-history
    baselines?
\end{itemize}

\subsection{Performance of the Non-Post-Quantum Instantiation (Halo2-based PCC)}
\label{sec:nonpq_eval}

We first evaluate the non-post-quantum instantiation of PCC built over
Halo2~\cite{halo2}. This part of the evaluation serves two purposes. First,
it compares PCC against \textbf{P-UC-DAC}, the closest prior practical
proof-based DAC baseline. Second, it contrasts two Halo2 realizations:
\textbf{ Halo2 Monolithic}, where the presentation circuit explicitly grows
with the delegation lineage, and \textbf{ Halo2 Recursive}, where lineage
validity is carried through the PCC state and the exported verifier-facing
presentation object is independent of the tested delegation depth. The key
question is whether the recursive PCC architecture removes the
depth-dependent verifier-facing cost observed in the monolithic baseline.

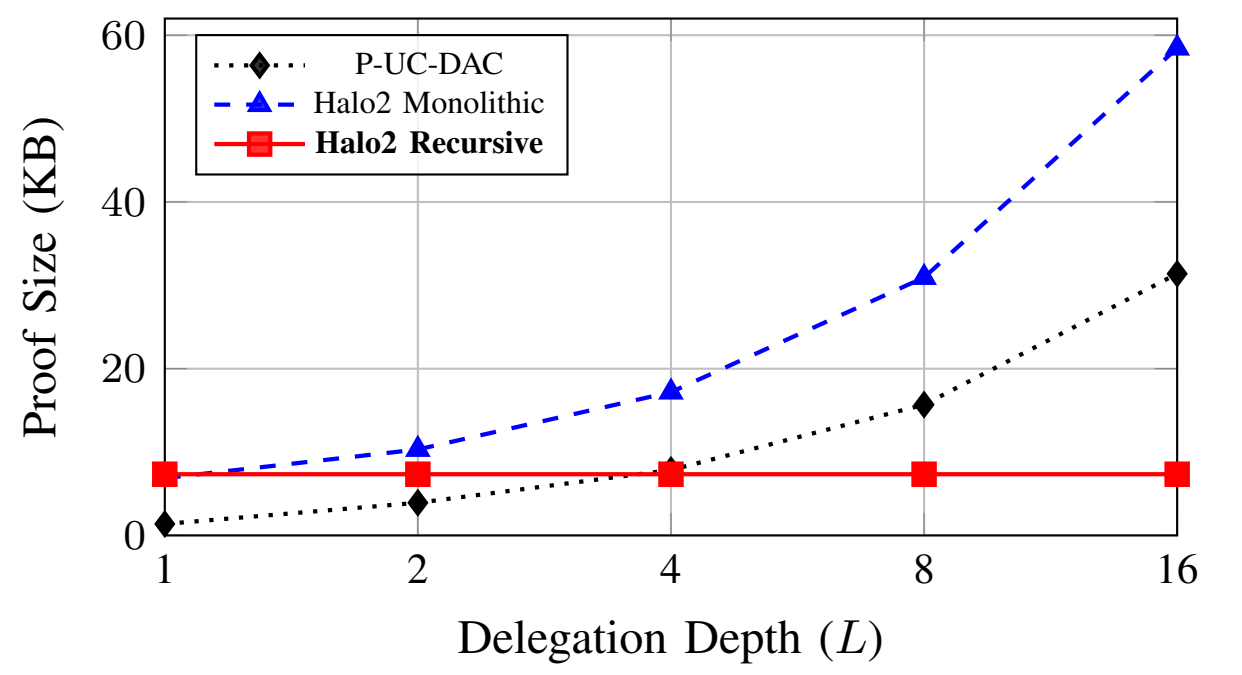
\begin{figure}[h]
    \centering
    \begin{tikzpicture}
        \begin{axis}[
            width=\linewidth,
            height=5cm,
            xlabel={Delegation Depth ($L$)},
            ylabel={Proof Size (KB)},
            xmin=1, xmax=16,
            ymin=0, ymax=62,
            xmode=log,
            log basis x={2},
            xtick={1, 2, 4, 8, 16},
            xticklabels={1, 2, 4, 8, 16},
            grid=major,
            legend pos=north west,
            legend style={
                nodes={scale=0.6, transform shape},
                fill=white,
                fill opacity=0.8,
                draw opacity=1,
                text opacity=1
            },
            label style={font=\small},
            tick label style={font=\footnotesize}
        ]

        % P-UC-DAC
        \addplot[color=black, mark=diamond*, dotted, thick, mark options={solid}] coordinates {
            (1, 1.38) (2, 3.92) (4, 7.85) (8, 15.69) (16, 31.39)
        };
        \addlegendentry{P-UC-DAC}

        % Halo2 monolithic
        \addplot[color=blue, mark=triangle*, dashed, thick, mark options={solid}] coordinates {
            (1, 6.876) (2, 10.314) (4, 17.190) (8, 30.942) (16, 58.446)
        };
        \addlegendentry{Halo2 Monolithic}

        % PCC Halo2 Recursive
        \addplot[color=red, mark=square*, thick, mark options={solid}] coordinates {
            (1, 7.344) (2, 7.344) (4, 7.344) (8, 7.344) (16, 7.344)
        };
        \addlegendentry{\textbf{Halo2 Recursive }}

        \end{axis}
    \end{tikzpicture}
    \caption{Proof size vs.\ delegation depth. The monolithic Halo2 baseline exposes depth-dependent verifier-facing proof material, whereas Recursive Halo2 carries lineage validity through the PCC state and exports a constant-size presentation object across the tested
depths.}
    \label{fig:proof_size_compare}
\end{figure}

\begin{figure}[h]
    \centering
    \begin{tikzpicture}
        \begin{axis}[
            width=\linewidth,
            height=5cm,
            xlabel={Delegation Depth ($L$)},
            ylabel={Verification Time (ms)},
            xmin=1, xmax=16,
            ymin=1, ymax=2500,
            xmode=log,
            log basis x={2},
            ymode=log,
            log basis y={10},
            xtick={1, 2, 4, 8, 16},
            xticklabels={1, 2, 4, 8, 16},
            grid=major,
            legend pos=north west,
            legend style={
                nodes={scale=0.6, transform shape},
                fill=white,
                fill opacity=0.8,
                draw opacity=1,
                text opacity=1
            },
            label style={font=\small},
            tick label style={font=\footnotesize}
        ]

        % P-UC-DAC verify
        \addplot[color=black, mark=diamond*, dotted, thick, mark options={solid}] coordinates {
            (1, 244) (2, 244) (4, 487) (8, 974) (16, 1949)
        };
        \addlegendentry{P-UC-DAC}

        % Halo2 Monolithic
        \addplot[color=blue, mark=triangle*, dashed, thick, mark options={solid}] coordinates {
            (1, 19.836) (2, 29.754) (4, 49.590) (8, 89.262) (16, 168.606)
        };
        \addlegendentry{Halo2 Monolithic}

        % PCC Halo2 Recursive
        \addplot[color=red, mark=square*, thick, mark options={solid}] coordinates {
            (1, 36.655) (2, 35.855) (4, 35.600) (8, 36.319) (16, 35.499)
        };
        \addlegendentry{\textbf{Halo2 Recursive}}

        \end{axis}
    \end{tikzpicture}
    \caption{Verification time vs.\ delegation depth. Monolithic Halo2 incurs depth-dependent verification cost because lineage-validation material grows with the delegation chain. Recursive Halo2 exports a depth-independent verifier-facing presentation object, yielding approximately constant verification time across tested depths.}
    \label{fig:verify_time_compare}
\end{figure}
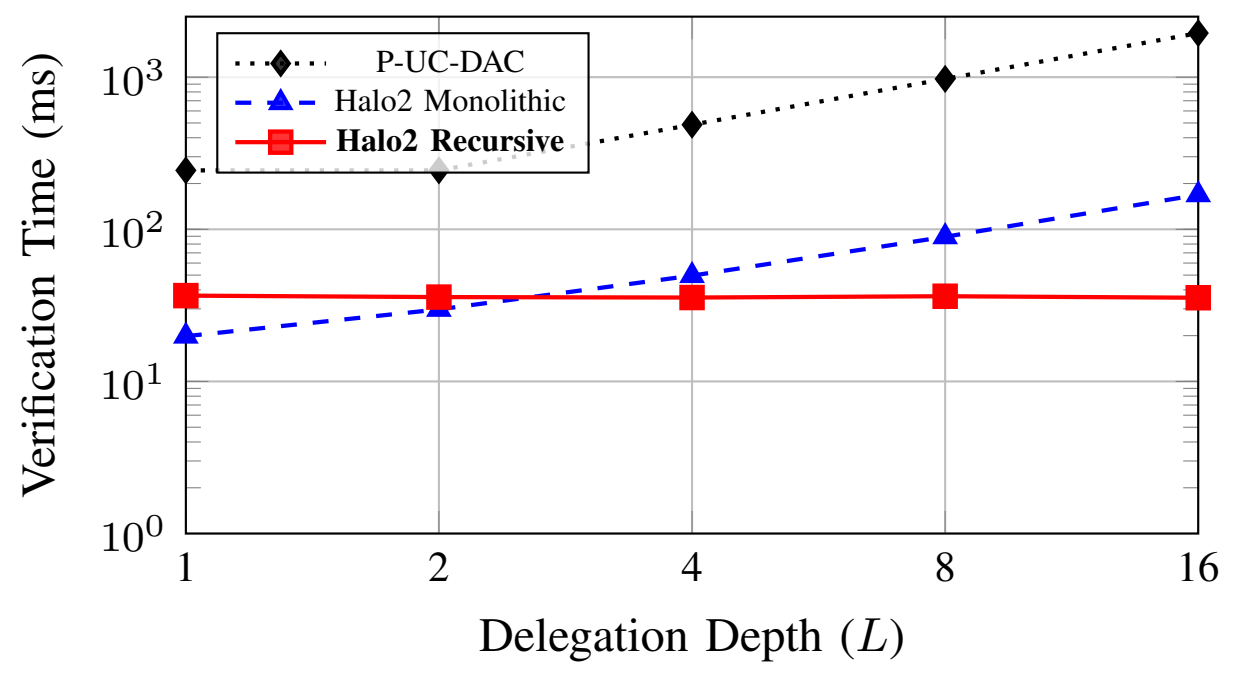

\subsubsection{Comparison}
\label{sec:haloq_baseline}

We begin with an external comparison against P-UC-DAC~\cite{camenisch2017practical}.
This baseline is the most relevant prior practical point of comparison for our non-post-quantum setting because, like PCC, it belongs to the proof-based DAC line rather than to the more recent signature-only constructions that explicitly avoid general zero-knowledge machinery.

Figure~\ref{fig:proof_size_compare} and Figure~\ref{fig:verify_time_compare} show the verifier-facing scaling behavior of the non-PQ instantiation. P-UC-DAC and monolithic Halo2 expose depth-dependent presentation material, causing proof size and verification latency to grow with L. Recursive Halo2 instead carries lineage validity through the PCC state and exports a constant-size verifier-facing proof: 7.344 KB and approximately 35–36 ms verification across all tested depths. Table III shows that the marginal online delegation cost is also stable across depths, confirming that each recursive update depends only on the previous proof and the current transition witness.

\textbf{Answer to RQ1.}
PCC removes the depth-dependent verifier-facing growth of the closest proof-based baseline and additionally hides the holder-side topology, trading a moderate constant recursive overhead for a proof interface whose size and verification cost are independent of the delegation depth.

\subsubsection{Delegation Cost}
\label{sec:haloq_runtime}
\looseness=-1
To complement the verifier-facing measurements, we also report the online cost of a single delegation step.
Table~\ref{tab:halo2_runtime} summarizes the delegation time per step together with the final verification cost and the final proof size.

\begin{table}[h]
    \centering
    \small
    \renewcommand{\arraystretch}{1.15}
    \caption{\textbf{Online delegation-step cost and final verifier-facing proof metrics for Halo2-based PCC.}}
    \label{tab:halo2_runtime}
    \begin{tabular}{cccc}
        \toprule
        \textbf{Depth} & \textbf{Delegation (s)} & \textbf{Verify (ms)} & \textbf{Proof (KB)} \\
        \midrule
        1  & 4.697 & 36.655 & 7.344 \\
        2  & 4.368 & 35.855 & 7.344 \\
        4  & 4.664 & 35.600 & 7.344 \\
        8  & 4.785 & 36.319 & 7.344 \\
        16 & 4.752 & 35.499 & 7.344 \\
        \bottomrule
    \end{tabular}
\end{table}

The \emph{delegation time per step} remains stable across all tested depths,
staying within a narrow range of 4.368--4.785\,s. Likewise, the final
verification cost varies only within 35.499--36.655\,ms across the tested
depths, consistent with the intended depth-independent verifier-facing
interface. This confirms that the online marginal cost of extending the
credential by one additional delegation step does not increase with prior
delegation depth. Together with the constant 7.344\,KB verifier-facing proof
and the approximately constant 35--36\,ms verification cost, this shows that
Halo2-based PCC supports incremental credential evolution without exposing a
depth-dependent proof interface to the verifier. 

We further verify that this depth-independence is not an artifact of a simple
disclosure policy: across simple ($\mathcal{L}_{\mathsf{sig}}$), range
($\mathcal{L}_{\mathsf{range}}$), and membership ($\mathcal{L}_{\mathsf{memb}}$)
presentation predicates implemented inside $\phi_{\mathsf{pre}}$, PCC keeps the
final verifier-facing proof at $7.344\,\mathrm{KB}$, so richer presentation logic
changes circuit content without reintroducing depth-dependent growth
(Table~\ref{tab:policy-complexity}, Appendix~\ref{app:policy}).

\textbf{Answer to RQ2.}
Under matched workloads, recursive Halo2 keeps a constant $7.344$\,KB proof and $35$ -$36$\,ms verification across all tested depths (per-step delegation stable at $4.368$--$4.785$\,s), whereas the monolithic variant grows with $L$; the depth-independence is therefore a property of the PCC architecture, not a Halo2 artifact.

\begin{table*}[t]
    \centering
    \small
    \renewcommand{\arraystretch}{1.15}
    \caption{\textbf{PQ architectural comparison at deployment-realistic delegation depths.}}
    \label{tab:comprehensive_comparison}
    \begin{adjustbox}{max width=\textwidth}
    \begin{tabular}{@{}c c | r r c | c c c | c c c@{}}
        \toprule
        \multirow{2}{*}{\textbf{Sec}} & \multirow{2}{*}{\textbf{Depth}} &
        \multicolumn{3}{c|}{\textbf{Circuit Complexity}} &
        \multicolumn{3}{c|}{\textbf{Prover Memory (MB)}} &
        \multicolumn{3}{c}{\textbf{Plonky2 Metrics}} \\
        & &
        \textbf{Base} & \textbf{Frac.} & \textbf{Plonky2} &
        \textbf{Base} & \textbf{Frac.} & \textbf{Plonky2} &
        \textbf{Prove/Step (s)} & \textbf{Verify (s)} & \textbf{Final Proof (KB)} \\
        \midrule

        % 100-bit
        \multirow{5}{*}{\textbf{100}}
          & 1  & 8,960$\dagger$   & 48,960$\dagger$ & 4,096$\clubsuit$ & 1.71  & 9.34 & 8.00  & 0.600 & 0.0029 & 125.2 \\
          & 2  & 17,920$\dagger$  & 48,960$\dagger$ & 8,192$\clubsuit$ & 3.42  & 9.34 & 16.00 & 0.600 & 0.0029 & 125.2 \\
          & 4  & 35,840$\dagger$  & 48,960$\dagger$ & 8,192$\clubsuit$ & 6.84  & 9.34 & 16.00 & 0.600 & 0.0029 & 125.2 \\
          & 8  & 71,680$\dagger$  & 48,960$\dagger$ & 8,192$\clubsuit$ & 13.67 & 9.34 & 16.00 & 0.600 & 0.0029 & 125.2 \\
          & 16 & 143,360$\dagger$ & 48,960$\dagger$ & 8,192$\clubsuit$ & 27.34 & 9.34 & 16.00 & 0.600 & 0.0029 & 125.2 \\
        \midrule

        % 128-bit
        \multirow{5}{*}{\textbf{128}}
          & 1  & 8,960$\dagger$   & 48,960$\dagger$ & 4,096$\clubsuit$ & 1.71  & 9.34 & 8.00  & 1.030 & 0.0038 & 157.9 \\
          & 2  & 17,920$\dagger$  & 48,960$\dagger$ & 8,192$\clubsuit$ & 3.42  & 9.34 & 16.00 & 1.030 & 0.0038 & 157.9 \\
          & 4  & 35,840$\dagger$  & 48,960$\dagger$ & 8,192$\clubsuit$ & 6.84  & 9.34 & 16.00 & 1.030 & 0.0038 & 157.9 \\
          & 8  & 71,680$\dagger$  & 48,960$\dagger$ & 8,192$\clubsuit$ & 13.67 & 9.34 & 16.00 & 1.030 & 0.0038 & 157.9 \\
          & 16 & 143,360$\dagger$ & 48,960$\dagger$ & 8,192$\clubsuit$ & 27.34 & 9.34 & 16.00 & 1.030 & 0.0038 & 157.9 \\
        \bottomrule
    \end{tabular}
    \end{adjustbox}

    \begin{minipage}{0.95\textwidth}
        \footnotesize
        \textbf{Notes:}
        \begin{itemize}[noitemsep,leftmargin=*]
            \item[$\dagger$] Complexity in \textbf{R1CS constraints}, used for the Fractal-based baselines.
            \item[$\clubsuit$] Complexity in \textbf{polynomial trace degree} ($N$), used for the Plonky2 instantiation.
            \item \textbf{Prove/Step} reports the cost of one additional recursive delegation step, not the total cost of constructing an entire depth-$L$ chain from scratch.
            \item We report only 100-bit and 128-bit settings because these are the security levels used in our concrete PQ evaluation.
        \end{itemize}
    \end{minipage}
\end{table*}
% This distinction is exactly the one required for long-lived delegated credential systems.
% In practice, the relevant online cost is the cost of the next transition, not a cumulative cost obtained by summing all prior steps.
% Halo2-based PCC achieves this online cost model while still exposing a depth-independent proof to the verifier at presentation time.
% \subsubsection{Expressiveness and Policy Complexity}
% \label{sec:haloq_policy}
% To check whether PCC's depth-independent presentation size is an artifact of a simple disclosure policy, we also evaluate richer presentation predicates implemented inside $\phi_{\mathsf{pre}}$. Table~\ref{tab:policy-complexity} should be read as a policy-complexity sanity check rather than a fully controlled cross-system microbenchmark: the PCC column reports measured Halo2 proof sizes, whereas the prior columns summarize representative proof-size contributions reported or implied by prior constructions. Across all tested policy classes, PCC keeps the final verifier-facing proof at 7.344 KB, indicating that richer presentation logic changes circuit content but does not reintroduce depth-dependent presentation growth.
\subsection{Performance of the Post-Quantum Instantiation (Plonky2-based PCC)}
\label{sec:pq_eval}
We next evaluate the scalability of our post-quantum (PQ) instantiation.
This part of the evaluation is architectural rather than lineage-comparative: unlike the non-post-quantum setting, there is no directly comparable prior DAC baseline built on recursive post-quantum proof machinery.
We therefore compare three PQ architectural variants:
\begin{itemize}[noitemsep,leftmargin=*]
    \item \textbf{Baseline (Fractal-Lin):} a non-recursive Fractal-style instantiation~\cite{chiesa2020fractal,libiop} in which signatures and state are aggregated linearly along the delegation chain;
    \item \textbf{Fractal (Fractal-Rec):} a recursive Fractal instantiation over R1CS, included as a functional recursive baseline for IVC-style delegation; and
    \item \textbf{Plonky2 (Ours):} our primary recursive PCC instantiation, using a Plonkish arithmetization over the Goldilocks field to optimize concrete proving efficiency.
\end{itemize}

We evaluate depths up to $L=16$, which already exceeds realistic deployment depths for transitive-authorization systems while still exercising the regime in which recursive closing should have stabilized.

Table~\ref{tab:comprehensive_comparison} compares non-recursive Fractal, recursive Fractal, and Plonky2-based PCC at 100-bit and 128-bit security. The non-recursive baseline grows linearly with L in constraints and memory, while both recursive designs remove this depth-dependent growth. In the 128-bit Plonky2 instantiation, PCC maintains a 157.9 KB final proof, 3.8 ms verification, and 1.03 s per recursive delegation step across all tested depths. These results support the main claim that PCC achieves depth-independent verifier-facing presentation with practical post-quantum recursive proving costs.
\textbf{Answer to RQ3.}
At 128-bit security, Plonky2-based PCC stays constant at $157.9$\,KB, $3.8$\,ms verification, and $1.03$\,s per delegation step across all teste depths, while the non-recursive Fracta baseline grows linearl ($143{,}360$ constraints, $27.34$\,MB at $L=16$); the cost is thus shifted to incremental delegation and remains practical at deployment-realistic depths.
% \par{Overall takeaway.}
% Overall, these results show that PCC achieves the intended recursive scaling behavior in the post-quantum setting.
% It avoids the linear-depth blowup of non-recursive designs, maintains a constant-size final verifier-facing proof object, and supports low-latency verification together with stable per-step recursive delegation cost at deployment-realistic depths.

\textbf{Summary and limitations.}
Overall, the experiments above support the claims tested by RQ1--RQ3: PCC improves the verifier-facing scaling profile over proof-based DAC baselines, the depth independence comes from recursive lineage carrying rather than a backend artifact, and the post-quantum instantiation remains practical at realistic delegation depths. The evaluation is limited to prototype implementations on a single workstation and delegation depths up to $L=16$; it does not measure network effects, side-channel leakage, revocation mechanisms, or multi-party deployment overhead. These issues are outside the current cryptographic threat model and require separate system-level evaluation.

% An anonymized repository with the implementation and evaluation scripts is available at \url{https://anonymous.4open.science/r/PCC-D3D2}.

% Table~\ref{tab:pq_complexity_measured} highlights the trade-off inherent to Post-Quantum recursion. At shallow depths ($L=1$), our PCC implementation incurs a significant "Recursion Overhead" (approx. $3\times$ to $5\times$ larger than the baseline) due to the fixed cost of the Fractal verifier circuit. 
% However, the linear growth of the baseline quickly overtakes this constant cost. For 80-bit security, our approach becomes more efficient at delegation depth \textbf{$L=4$}. For higher security levels (100/128-bit), where the verifier circuit is larger, the cross-over point occurs at \textbf{$L=8$}. By $L=16$, PCC offers a substantial \textbf{$3\times$ to $5\times$ reduction} in circuit complexity.

% A critical advantage of our architecture is evident in the memory analysis. The baseline approach requires linear memory growth, reaching $\approx 27$ MB at $L=16$. In contrast, our PCC instantiation maintains a strictly constant memory footprint ($\approx 5.5$ MB for 80-bit, $\approx 9.3$ MB for 128-bit) regardless of depth. This constant-memory property is vital for deploying PQ-secure credentials on hardware-constrained devices, ensuring that deep delegation chains do not trigger Out-Of-Memory (OOM) failures.

\subsection{Presentation Latency and Memory under Matched FRI Configurations}
\label{sec:matched-fri-eval}

\begin{figure*}[t]
    \centering
    \includegraphics[width=.86\textwidth]{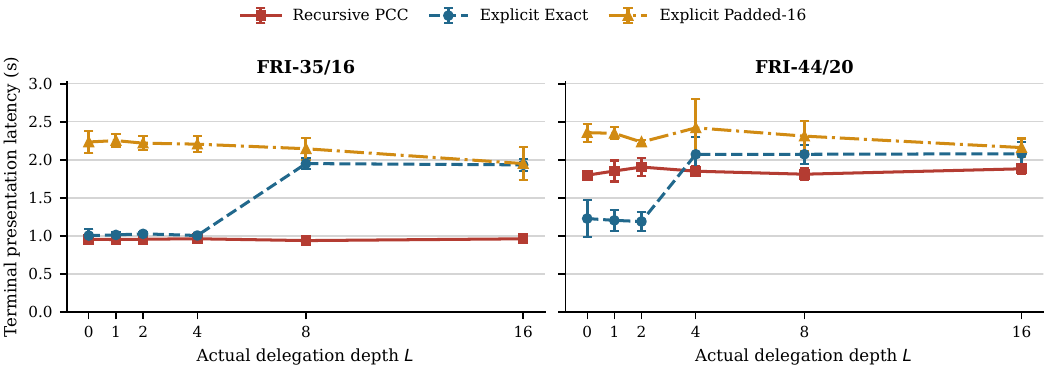}\\[-1pt]
    {\footnotesize\textbf{(a)} Terminal presentation latency.}\\[2pt]
    \includegraphics[width=.86\textwidth]{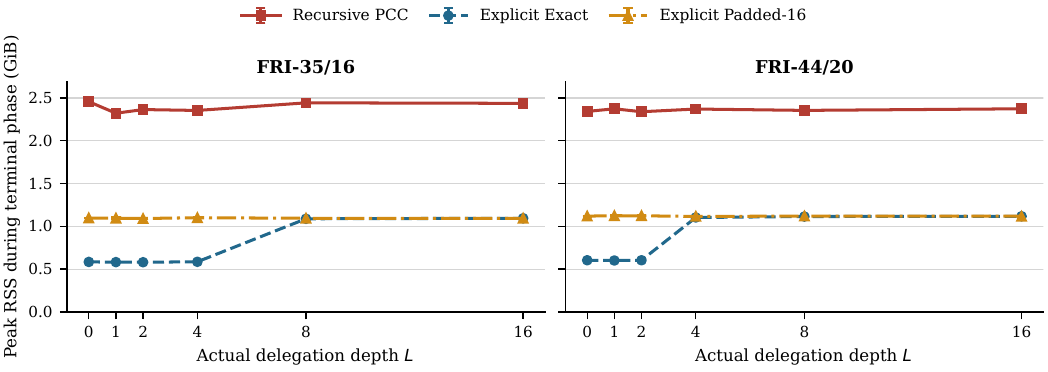}\\[-1pt]
    {\footnotesize\textbf{(b)} Absolute peak terminal resident memory (RSS).}
    \caption{Matched FRI measurements across actual delegation depths
    (16 threads, five independent processes per cell). The latency bars
    are two-sided Student-$t$ 95\% confidence intervals for the mean.}
    \label{fig:matched-fri-terminal}
\end{figure*}

To isolate the effect of the credential representation from the backend,
we compare recursive PCC with two explicit-history baselines in a
separate FRI realization with a dedicated presentation circuit.
Explicit-Exact checks the actual chain, whereas Explicit-Padded-16
checks a fixed 16-slot relation including inactive slots. Within each
profile, all systems use the same Goldilocks/Poseidon stack, hash-based
authorization, eight-bit policy and public showing inputs.
We run five independent processes per system, profile and depth
$L\in\{0,1,2,4,8,16\}$ (180 runs), at 16 threads on an AMD Ryzen~7
9800X3D with 31.15~GiB of RAM under Windows. Presentation timing excludes
circuit construction, issuance and delegation. The profiles denote FRI
query and proof-of-work settings, not certified security levels.

Figure~\ref{fig:matched-fri-terminal}(a) shows that PCC presentation
latency is nearly flat in depth: 0.939--0.963~s for FRI-35/16 and
1.799--1.906~s for FRI-44/20. Across the tested depths, PCC's mean
latency is 50.7--57.7\% and 12.9--23.7\% lower than Padded-16,
respectively. Exact is faster at the three shallowest depths under
FRI-44/20. Figure~\ref{fig:matched-fri-terminal}(b) shows the cost of
the recursive state: PCC uses about $2.1$--$2.2\times$ the peak RSS of
Padded-16. These are separate process-level measurements, not direct
comparisons to the timings or memory figures in
Table~\ref{tab:comprehensive_comparison}.

\textbf{Answer to RQ4.}
PCC's presentation latency and peak resident memory stay nearly flat
across the measured depths. Its mean presentation latency is lower than
Padded-16's in both profiles, but its peak resident memory is higher;
Explicit-Exact is faster at shallow depths under FRI-44/20.

\section{Related Work}
\label{sec:related}
% We focus on recursive proof systems and proof-carrying mechanisms; Section~\ref{sec:intro} covers the DAC design space and anonymity trade-offs. Incrementally Verifiable Computation (IVC)~\cite{valiant2008ivc} and its accumulation- and folding-based realizations~\cite{bowe2019halo,kothapalli2022nova}, generalized to Non-Uniform IVC for heterogeneous steps~\cite{kothapalli2022supernova,kothapalli2024hypernova}, provide our recursive substrate, which PCC drives with separate issuance, delegation, and presentation predicates; none of these define a DAC abstraction, delegated authorization, or holder-visible topology leakage. Proof-carrying data (PCD)~\cite{chiesa2010pcd,bunz2020pcd} attaches proofs to data in distributed computations, whereas PCC's carried proof certifies credential lineage and authorization while hiding remote topology from the terminal holder. 

% Our concrete backend builds on transparent post-quantum systems: STARK/FRI integrity~\cite{ben2018stark,ben2018fri}, its recursive form in Fractal~\cite{chiesa2020fractal}, Plonky2's Plonkish/Goldilocks efficiency~\cite{plonky2}, and Poseidon hashing~\cite{poseidon}. Derecho~\cite{beal2024derecho} applies proof-carrying disclosures but addresses neither delegated authorization nor the holder-side topology hiding captured by EDA.

We focus on recursive proof systems and proof-carrying mechanisms most directly related to PCC; Section~\ref{sec:intro} discusses the DAC design space and anonymity trade-offs.

Incrementally Verifiable Computation (IVC) was formalized by Valiant~\cite{valiant2008ivc}. Halo~\cite{bowe2019halo} introduced accumulation-based recursion without trusted setup, and Nova~\cite{kothapalli2022nova} improved recursive proving through folding. SuperNova~\cite{kothapalli2022supernova} generalized this line to Non-Uniform IVC (NIVC) for heterogeneous step functions, while HyperNova~\cite{kothapalli2024hypernova} further improves folding-based recursion. PCC uses this abstraction for separate issuance, delegation, and presentation predicates. These systems provide the recursive-proof substrate, but do not define a DAC abstraction, address delegated authorization, or model holder-visible topology leakage.

PCC is also related to proof-carrying data (PCD): Chiesa and Tromer introduced PCD for attaching proofs to data flowing through distributed computations~\cite{chiesa2010pcd}, and subsequent work connected PCD to accumulation schemes~\cite{bunz2020pcd}. PCC differs in both object and privacy goal: its carried proof certifies credential lineage and authorization validity while hiding remote delegation topology from the terminal holder.

Transparent and post-quantum proof systems form our concrete backend. STARKs and FRI-based proof systems provide transparent post-quantum computational integrity~\cite{ben2018stark,ben2018fri}; Fractal applies this line to recursion~\cite{chiesa2020fractal}, and Plonky2 improves concrete efficiency via Plonkish arithmetization over the Goldilocks field~\cite{plonky2}. We use Poseidon for arithmetization-friendly hashing~\cite{poseidon}. Derecho~\cite{beal2024derecho} also uses proof-carrying techniques for privacy-preserving disclosures, but it does not address delegated authorization or the holder-side topology-hiding requirement captured by EDA.
\section{Conclusion and Future Work}
\label{sec:conclusion}
We introduced Enhanced Delegation Anonymity (EDA), which pairs verifier-side anonymity with holder-side topology hiding, and Proof Carrying Credentials (PCC), a recursive DAC framework that compiles delegation lineage into a depth-independent proof-carrying state without sacrificing restricted delegation or expressive presentation. Our Halo2 and Plonky2 instantiations achieve this at practical proving cost, including a post-quantum-compatible QROM instantiation. Future work includes revocation, threshold delegation, and broader recursive backends.

% \section*{Acknowledgments}
% OpenAI Codex assisted in editing the topology-hiding motivation and
% in revising the security definitions and proof arguments.

\bibliographystyle{IEEEtran}
\bibliography{bib}
\appendices
\section{Supplementary Protocol Details}
\subsection{NIVC: Full Syntax and Security Properties}
\label{app:nivc}
\begin{definition}[PCC-Compatible Non-Uniform Incrementally Verifiable Computation]
\label{def:nivc}
A PCC-compatible Non-Uniform Incrementally Verifiable Computation scheme
for a set of $k$ polynomial-time, deterministic state-transition
verification functions
$
    \F = \{\F_1, \F_2, \dots, \F_k\}
$
is a tuple of four polynomial-time algorithms
$
    (\setup, \keygen, \prove, \cverify)
$
with the following syntax:
We write $\rho$ for public instance data shared by all transition predicates; in PCC, $\rho$ includes the authenticated root anchor $\RootAnchor$.
\begin{itemize}[noitemsep,leftmargin=*, align=left]
    \item $\setup(1^\lambda) \rightarrow \pp$:
    A probabilistic algorithm that takes a security parameter $\lambda$
    and outputs public parameters $\pp$. These parameters are assumed to
    be implicitly available to all other algorithms.
    
    \item $\keygen(\pp, \F) \rightarrow (\proverk, \verifierk)$:
    A deterministic algorithm that takes the public parameters $\pp$ and
    a description of the transition-function family $\F$ to produce a
    prover key $\proverk$ and a verifier key $\verifierk$.
     
    \item
    $\prove(\proverk, j, \rho, z_0, z_i, z_{i+1}, w_{i+1}, \pi_i)
    \rightarrow \pi_{i+1}$:
    A probabilistic algorithm for an incremental step. It takes the prover
    key $\proverk$, a function selector $j \in \{1,\dots,k\}$, the public instance data $\rho$, the initial
    state $z_0$, the current state $z_i$, the next state $z_{i+1}$, a
    non-deterministic transition witness $w_{i+1}$, and the previous
    recursive proof $\pi_i$. It outputs an updated proof $\pi_{i+1}$ of
    constant size, attesting that the previous recursive computation is
    valid and that the current transition satisfies
    $
        \F_j(\rho,z_i,z_{i+1},w_{i+1})=1.
    $
    For the base step, $\pi_0$ is taken to be a distinguished empty proof $\bot$, which only initializes recursion and does not authenticate the initial credential state. In PCC, root authenticity is enforced by the $\phi_{\mathsf{issue}}$ transition using the root verification key contained in $\rho$.
    \item
    $\cverify(\verifierk, \rho, z_0, z_n, \pi_n, \tau_{\mathsf{pub}})
    \rightarrow \{0,1\}$:
    A deterministic algorithm that takes the verifier key $\verifierk$,
    the public instance data $\rho$, the initial public state $z_0$, the final public state $z_n$, the final
    recursive proof $\pi_n$, and a public terminal condition
    $\tau_{\mathsf{pub}}$. It outputs $1$ if $\pi_n$ certifies the
    existence of a valid recursive execution from $z_0$ to $z_n$ satisfying
    $\tau_{\mathsf{pub}}$, and outputs $0$ otherwise.
\end{itemize}
\end{definition}

The public recursive statement is
$x=(\rho,z_{\mathsf{root}},z_{\mathsf{out}},\tau_{\mathsf{pub}})$.
The terminal conditions are typed: $\tau_{\mathsf{cred}}$ denotes an issue
followed by zero or more delegations; $\tau_{\mathsf{pre}}$ denotes such a
credential trace followed by one presentation. These are conditions on the
entire execution, not just on its last selector. In particular, issue is
the only base transition, and a presentation proof is not a credential
predecessor. The actual length, non-terminal selectors, intermediate
states, and witnesses are not public inputs.

\begin{definition}[PCC-compatible proof security]
\label{def:pcc-proof-interface}
In addition to completeness for the typed execution language
(Definition~\ref{def:valid-lineage}), the NIVC interface requires the
following properties in a common setup and oracle environment.
\begin{itemize}[leftmargin=*,noitemsep]
\item \emph{Adaptive auxiliary-input zero knowledge.} One stateful
simulator produces proof objects for true statements $x$, without the
private penultimate state, selector history, or witnesses. Replacing honest
proofs by these simulations remains indistinguishable in the presence of
the games' permitted auxiliary data and oracle responses, including
honestly produced credential proofs when simulating presentations. For
holder privacy the simulated credential must also support the public
presentation-proving operation using the fixed local witness.
\item \emph{Fresh-statement extraction after simulated showings.} For an
accepting output $(x^*,\pi^*)$ with $x^*$ outside the log of simulated
statements, an extractor obtains a finite typed lineage with its
transition witnesses for that output, except with negligible failure.
This must hold in the presence of genuinely produced credential proofs
and permitted system-oracle responses. The extractor may use the backend's
simulation/extraction state and retained honest credential witnesses.
It must operate on the same accepting output and signer transcript,
without the external signing challenge secret, resetting external state,
or adding signing queries.
\end{itemize}
The trace has polynomial length, and the extraction work is expected
polynomial in the security parameter and the experiment's polynomial
trace horizon, including locally generated updates. This horizon is a
proof parameter, not a serialized depth input. These are explicit recursive composition
requirements; ordinary soundness or one-step knowledge soundness does
not by itself give them~\cite{chiesa2024security,Pcd2025}.
\end{definition}
\subsection{Primitive Details for the Post-Quantum Instantiation}
\label{app:pq-primitives}
\spar{Field Arithmetic}
We choose the 64-bit Goldilocks prime field $\Fp$ with modulus $p = 2^{64} - 2^{32} + 1$. 
This specific prime is chosen fundamentally for its hardware alignment: unlike larger fields that require multi-precision arithmetic, Goldilocks operations fit natively within 64-bit CPU registers and support aggressive SIMD vectorization, thereby maximizing the prover's arithmetic throughput. 
Furthermore, the modulus satisfies $p \equiv 1 \pmod{2^{32}}$, ensuring high 2-adicity to support the efficient Number Theoretic Transforms (NTT) required for polynomial commitments.
The algebraic circuit is defined over $\Fp$ for efficiency, while FRI
challenges are sampled from the quadratic extension $\Fp^2$. The size of
that challenge space is not a bound on the protocol's soundness error or
an end-to-end post-quantum security level.
\begin{table}[t]
\centering
\caption{Primitive stack for the post-quantum-compatible PCC instantiation.}
\label{tab:pq-primitives}
\footnotesize
\setlength{\tabcolsep}{4pt}
\renewcommand{\arraystretch}{1.3}
\rowcolors{2}{white}{rowalt}
\begin{adjustbox}{max width=\columnwidth}
\begin{tabular}{@{} >{\raggedright\arraybackslash}p{0.20\columnwidth}
                    >{\raggedright\arraybackslash}p{0.30\columnwidth}
                    >{\raggedright\arraybackslash}p{0.40\columnwidth} @{}}
\toprule
\textbf{Component} & \textbf{Instantiation} & \textbf{Role / Parameter Choice} \\
\midrule
\multicolumn{3}{@{}l}{\textit{Algebraic layer}}\\
\midrule
Field arithmetic
  & Goldilocks $\Fp$, $p=2^{64}-2^{32}+1$; FRI over $\Fp^{2}$
  & 64-bit native arithmetic; high 2-adicity; extension-field challenges. \\
Hash function
  & Poseidon / Goldilocks; capacity $c=4$, width $w=12$
  & State binding, Merkle authentication, FS hashing in-circuit. \\
\midrule
\multicolumn{3}{@{}l}{\textit{Cryptographic primitives}}\\
\midrule
Issuance and delegation signature
  & Poseidon-based hash-based sig.\ (SPHINCS\textsuperscript{+}-style)
  & PQ root issuance and delegation authorization; circuit-friendly verification. \\
Recursive proof system
  & Plonky2-style FRI recursion (Fractal-compatible)
  & Transparent recursion; no pairings; no trusted setup. \\
\midrule
\multicolumn{3}{@{}l}{\textit{Protocol interfaces}}\\
\midrule
Oracle domains
  & $H_{\mathsf{state}}$, $H_{\mathsf{sig}}$, $H_{\mathsf{FS}}$, $H_{\mathsf{pre}}$
  & Domain separation for state, signatures, FS, presentation binding. \\
\bottomrule
\end{tabular}
\end{adjustbox}
\end{table}
\spar{Arithmetization-Friendly Hash Function} 
A critical component of our system is the domain-separated hash layer
implementing $H_{\mathsf{state}}$, $H_{\mathsf{sig}}$, $H_{\mathsf{pre}}$,
and $H_{\mathsf{FS}}$.
We instantiate $\phash$ using Poseidon hash~\cite{poseidon}, which is
designed for low algebraic complexity.
\begin{itemize}[noitemsep,leftmargin=*]
    \item \textbf{Configuration:} We instantiate the Poseidon hash function over the 64-bit Goldilocks field. 
    We configure the sponge with capacity $c=4$ field elements and state
    width $w=12$. These are parameters, not an estimate of concrete
    classical or quantum attack cost.
    \item \textbf{Efficiency:} Unlike SHA-3, which relies on bitwise operations that incur prohibitive overhead in arithmetic circuits, Poseidon is designed natively for algebraic execution. 
    Furthermore, our use of the 64-bit Goldilocks field allows for vectorized arithmetic (e.g., AVX-512) on modern CPUs. 
    This design significantly reduces the number of constraints per round and enables orders-of-magnitude faster proof generation compared to schemes operating over 256-bit prime fields.
\end{itemize}
\spar{Post-Quantum Signature Scheme ($\Sigma$)}
To instantiate the signature scheme $\Sigma$ used for root issuance and
delegation authorization, we use a Plonky2-native hash-based signature
construction.
Structurally analogous to SPHINCS+~\cite{SPHINCS}, our construction is tailored for arithmetic circuit efficiency by exclusively employing the Poseidon hash function over the Goldilocks field. 
In this scheme, the verification key is defined as the root of a Merkle tree, where leaves correspond to one-time signing keys (instantiated via Poseidon-based W-OTS). 
Consequently, a valid signature consists of a one-time signature and the Merkle authentication path linking the leaf to the root. 
This design allows the verification logic to be realized as a highly efficient native gadget within the recursive circuit, incurring negligible overhead compared to generic post-quantum signatures.
The analysis assumes EUF-CMA of the signing implementation, including
safe management of one-time signing leaves. The structural analogy is
not an identification with standardized SPHINCS\textsuperscript{+} or a
proof of stateless security.

\spar{Post-Quantum NIVC}
The realization of post-quantum NIVC primarily relies on recursive arguments constructed from the Fast Reed-Solomon Interactive Oracle Proofs of Proximity (FRI) protocol. Fractal~\cite{chiesa2020fractal} pioneered this direction by instantiating the recursive accumulation paradigm over Rank-1 Constraint Systems (R1CS), demonstrating for the first time that recursive composition is feasible without relying on quantum-vulnerable bilinear pairings. 
Building upon this foundation, Plonky2~\cite{plonky2} significantly reduced the recursion overhead by adopting Plonkish arithmetization and operating over the 64-bit Goldilocks field, which enables highly efficient, vectorized field arithmetic on modern CPUs. 
A defining advantage of these FRI-based constructions is their use of
hash functions without a trusted setup. Their post-quantum proof properties
remain conditional on the assumptions in Appendix~\ref{app:qrom-model}.
While our PCC allows for flexible instantiation with any FRI-based recursive system, we prioritize Plonky2 in our primary construction to leverage its specialized circuit optimizations. 
However, to provide a comprehensive comparative analysis of circuit complexity and scalability, we implement and evaluate both the Fractal-based and Plonky2-based instantiations in Section~\ref{sec:eval}.
\subsection{Detailed construction of the PQDAC Protocol}
\label{app:pqdac-protocol}
Figure~\ref{fig:pqdac_protocol} gives the full algorithmic specification of
$\Pi_{\mathsf{PQDAC}}$, the post-quantum instantiation summarized in
Section~\ref{sec:protocol}. It makes explicit the witness and argument
construction that the wrapper $\pccderive$ encapsulates at each phase:
$\textsf{Setup}$ compiles $\Phi$ into the proving and verification keys;
$\textsf{KeyGen}$ derives user keys as post-quantum signature keypairs; and
$\textsf{Issue}$, $\textsf{Delegate}$, and $\textsf{Present}$ are single
$\pccderive$ calls in issuance, delegation, and presentation mode,
respectively, each extending the carried token by one transition.
$\textsf{Verify}$ reduces to a single $\pccverify$ check against the
authenticated $\RootAnchor$. All cryptographic logic is inherited from the
generic $\Pi_{\mathsf{PCC}}$ (Definition~\ref{def:pcc_scheme}); the only
instantiation-specific choices are the primitives of
Table~\ref{tab:pq-primitives}.
\subsection{Presentation-Policy Complexity}
\label{app:policy}
Table~\ref{tab:policy-complexity} reports the final verifier-facing proof size
under presentation predicates of increasing complexity. It is a policy-complexity
sanity check rather than a fully controlled cross-system microbenchmark: the PCC
column reports measured Halo2 proof sizes, whereas the prior columns summarize
representative proof-size contributions reported or implied by prior constructions.
Across all tested policy classes, PCC keeps the final proof at $7.344\,\mathrm{KB}$.

\begin{table}[ht]
    \centering
    \small
    \setlength{\tabcolsep}{2pt}
    \renewcommand{\arraystretch}{1.2}
    \caption{Impact of presentation-policy complexity on final proof size.
    Prior columns are literature-based estimates illustrating qualitative
    scaling; PCC reports measured Halo2 final proof size.}
    \label{tab:policy-complexity}
    \begin{adjustbox}{max width=\columnwidth}
        \begin{tabular}{@{} l c c c c @{}}
            \toprule
            \textbf{Policy} & \textbf{Logic} & \textbf{P-UC-DAC~\cite{camenisch2017practical}} & \textbf{Mercurial~\cite{mercurial2024}} & \textbf{Ours} \\
            \midrule
            \textbf{Simple}  & $\mathcal{L}_{\mathsf{sig}}$   & $\approx 0.5$ KB          & $\approx 0.2$ KB         & $7.344$ KB \\
            \textbf{Medium}  & $\mathcal{L}_{\mathsf{range}}$ & $\approx 1.5$ KB\textsuperscript{a} & $\approx 1.5$ KB\textsuperscript{a} & $7.344$ KB \\
            \textbf{Complex} & $\mathcal{L}_{\mathsf{memb}}$  & $\approx 20$ KB\textsuperscript{b}  & $\approx 20$ KB\textsuperscript{b}  & $7.344$ KB \\
            \bottomrule
        \end{tabular}
    \end{adjustbox}
    \par\smallskip
    \begin{minipage}{\columnwidth}\raggedright\footnotesize
        \textsuperscript{a} Range-proof contribution to the presentation size.\\
        \textsuperscript{b} Set-membership-proof contribution to the presentation size.
    \end{minipage}
\end{table}

\begin{figure*}[t!]
\centering
\fbox{%
\begin{minipage}{\dimexpr\textwidth-2\fboxsep-2\fboxrule\relax}
    % --- 标题 ---
    \centering
    \textbf{\large The PQDAC Protocol}
    \par
    \hrule height 1pt
    
    % --- 双栏开始 ---
    \begin{minipage}[t]{0.48\textwidth}
        % ================= LEFT COLUMN =================
        
        % --- Setup ---
        \textbf{\textsf{Setup}}$(1^\lambda)\to (\pp,\proverk,\verifierk)$:
        \begin{enumerate}[leftmargin=*, nosep, label=\arabic*.]
            \item Generate generic public parameters for the underlying PCC scheme:
            $\pp \leftarrow \pccsetup(1^\lambda)$.
            \item Define
            $\Phi=\{\phi_{\mathsf{issue}},\phi_{\mathsf{del}},\phi_{\mathsf{pre}}\}$
            as the recursive relation logic for issuance, delegation, and presentation.
            \item Compile the proving and verification keys using the PCC indexer:
            $(\proverk,\verifierk)\leftarrow \pcckeygen(\pp)$.
            \item Return $( \pp,\proverk,\verifierk )$.
        \end{enumerate}
        \par \hrule  % 分割线

       % --- KeyGen ---
        \textbf{\textsf{KeyGen}$(\pp) \to (\usk, \upk)$:}
        \begin{enumerate}[leftmargin=*, nosep, label=\arabic*.]
            \item Parse public parameters $\pp$ to retrieve the security parameter $\lambda$ and the underlying finite field $\mathbb{F}$.
            \item Run the standard key generation algorithm for the post-quantum signature component:\\
            $(\sk_{\mathsf{sig}}, \pk_{\mathsf{sig}}) \leftarrow \Sigma\mathsf{.KeyGen}(1^\lambda)$.
            % \item Sample a uniform random blinding factor (trapdoor) for future attribute commitments: $\tau \leftarrow \mathbb{F}^*$.
            % \item \textit{// Assemble keys ensuring binding to the identity}
            \item Set the user secret key: $\usk := \sk_{\mathsf{sig}}$.
            \item Set the user public key: $\upk := \pk_{\mathsf{sig}}$.
            \item Return $(\usk, \upk)$.
        \end{enumerate}
        \par \hrule % 分割线

       % --- Issue ---
        \textbf{\textsf{Issue}$(\RootAnchor,\upk_0,A_0;\isk_{\mathsf{root}})\to\pcred_0$:}
        \begin{enumerate}[leftmargin=*, nosep, label=\arabic*.]
            \item Set $\rho:=\RootAnchor$ and
            $z_{\mathsf{root}}:=H_{\mathsf{state}}(\mathsf{tag}_{\mathsf{issue}},\rho)$.
            \item Define issuance arguments:     
            
            $\mathsf{args}_{\mathsf{iss}}:=(\upk_0,A_0,\mathsf{meta}_{\mathsf{iss}}).$
            \item Invoke the PCC wrapper in issuance mode:
            \[
            \token_0 \leftarrow
            \pccderive(\rho,z_{\mathsf{root}},\bot,
            \mathsf{args}_{\mathsf{iss}};\isk_{\mathsf{root}}).
            \]
            \item Construct the initial credential:
            \[\pcred_0 := (\rho,\token_0).\]
            \item Return $\pcred_0$.
        \end{enumerate}
        
    \end{minipage}
    \hfill\vrule\hfill % 中间垂直分割线 (可选,如果觉得太挤可以去掉 \vrule 只留 \hfill)
    \begin{minipage}[t]{0.48\textwidth}
        % ================= RIGHT COLUMN =================

        % --- Delegate ---
    \textbf{\textsf{Delegate}$(\pcred_i,\upk_{i+1},A_{i+1};\usk_i)\to\pcred_{i+1}$:}
        \begin{enumerate}[leftmargin=*, nosep, label=\arabic*.]
            \item Parse input credential $\pcred_i$ as $(\rho,\token_i)$ and set
            $z_{\mathsf{root}}:=H_{\mathsf{state}}(\mathsf{tag}_{\mathsf{issue}},\rho)$.
            \item Define
            $\mathsf{args}_{i+1}:=(\upk_{i+1},A_{i+1},\mathsf{meta}_{\mathsf{del}})$.
            Invoke the PCC recursive wrapper to delegate authority to $U_{i+1}$:
            \[
            \begin{aligned}
            \token_{i+1} &\leftarrow
            \pccderive(\rho,z_{\mathsf{root}},\token_i,\\
            &\qquad\mathsf{args}_{i+1};\usk_i).
            \end{aligned}
            \]
            \item This step recursively validates $\token_i$ and enforces the delegation predicate $\phi_{\mathsf{del}}$.
            \item Construct the new delegated credential:\\
            $\pcred_{i+1}:=(\rho,\token_{i+1})$.
            \item Return $\pcred_{i+1}$.
        \end{enumerate}
        \par \hrule % 分割线

        % --- Present ---
        \textbf{\textsf{Present}$(\pcred_n, \usk_n, \policy, m,A_{\mathsf{pre}}) \to \Pi$:}
        \begin{enumerate}[leftmargin=*, nosep, label=\arabic*.]
            \item Set
            $z_{\mathsf{root}}:=H_{\mathsf{state}}(\mathsf{tag}_{\mathsf{issue}},\rho)$ and parse $\pcred_n$ as $(\rho,\token_n)$.
            \item Define presentation context $\mathsf{ctx}:=(m,\policy)$ and
            $\mathsf{args}_{\mathsf{pre}}:=(\mathsf{ctx},A_{\mathsf{pre}})$.
            \item Invoke the PCC wrapper in presentation mode:
              \[
                \begin{aligned}
                \token_{\mathsf{pre}} &\leftarrow
                \pccderive(\rho,z_{\mathsf{root}},\token_n,\\
                &\qquad\mathsf{args}_{\mathsf{pre}};\usk_n).
                \end{aligned}
            \]
            \item Parse
            $\token_{\mathsf{pre}}=(z_{\mathsf{pre}},\pi_{\mathsf{pre}},\mathsf{ctx},A_{\mathsf{pre}}).$
            \item Return presentation
            $\Pi := (z_{\mathsf{pre}}, \pi_{\mathsf{pre}}, m, \policy,A_{\mathsf{pre}})$.
        \end{enumerate}
        \par \hrule % 分割线

        % --- Verify ---
        \textbf{\textsf{Verify}$(\verifierk, \RootAnchor, \Pi) \to \{0, 1\}$:}
        \begin{enumerate}[leftmargin=*, nosep, label=\arabic*.]
            \item Parse $\Pi$ as
            $(z_{\mathsf{pre}}, \pi_{\mathsf{pre}}, m, \policy,A_{\mathsf{pre}})$.
            \item Set $\rho:=\RootAnchor$, reconstruct
            $\mathsf{ctx}:=(m,\policy)$, and form
            $\token_{\mathsf{pub}}:=(z_{\mathsf{pre}},\pi_{\mathsf{pre}},\mathsf{ctx},A_{\mathsf{pre}}).$
            \item Invoke the PCC verifier relative to the authenticated root anchor:
            $b \leftarrow \pccverify(\verifierk,\rho,\token_{\mathsf{pub}})$.
            \item Return bit $b$.
        \end{enumerate}
        
    \end{minipage}
\end{minipage}%
}
\caption{Detailed construction of the PQDAC Protocol instantiated over the generic PCC. The left column describes the system setup and issuance phases, while the right column details the delegation, presentation, and verification processes.}
\label{fig:pqdac_protocol}
\end{figure*}
\section{PQDAC Security}
\label{appendix}

\subsection{Assumptions and QROM Model}
\label{app:qrom-model}
The proofs below require the typed NIVC interface in
Definition~\ref{def:pcc-proof-interface}, EUF-CMA security of $\Sigma$,
binding state commitments, and collision resistance of the hashes on
canonically encoded signed and public statements. In the QROM version the
attacker has quantum access to the modeled hash interfaces, but the signing and DAC service
queries are classical. We distinguish the domains $H_{\mathsf{state}}$,
$H_{\mathsf{sig}}$, $H_{\mathsf{FS}}$, and $H_{\mathsf{pre}}$ for the state
binding, signature operations, Fiat--Shamir transcript, and presentation
binding, respectively. The simulated and real proofs must be compatible
with the same setup and oracle environment, including the interactions
with honestly produced credential proofs and signing responses. Any
Fiat--Shamir programming and its loss are part of the assumed proof
interface, not consequences of domain separation.

Modeling a transcript hash as a quantum-accessible random oracle does not
automatically provide a circuit for evaluating an ideal oracle inside a
recursive proof; arithmetic random-oracle models distinguish these
interfaces~\cite{chen2023arom}. The role of in-circuit concrete Poseidon must be
justified separately in any reduction to an ideal-oracle model. Neither
ordinary FRI soundness nor the extension-field and sponge parameters
alone establish the extraction and zero-knowledge assumptions for the
evaluated backend. The conclusion below is conditional and does not
assign a concrete quantum security level.

\subsection{Delegation Unforgeability}
\label{app:pqdac-unforge}

\begin{definition}[Usable terminal signing witness]
\label{def:pcc-key-relation}
When the presentation predicate accepts a secret-key witness $\usk$ as
corresponding to the opening's committed key $\upk$, that witness must
enable the reduction to sign a fresh digest that verifies under $\upk$.
For an ordinary signature key pair this is the key relation with signing
correctness. If the signing interface requires additional state, the
terminal witness or the relation must permit recovery of sufficient usable
state without the honest holder's secret. A check on a cached public-key
field alone would not suffice.
\end{definition}

\begin{proof}[Proof of Theorem~\ref{thm:pqdac-unforge}]
\leavevmode\par
Use the honest-signing ledger of Definition~\ref{def_unforgeability}.
First replace each successful honest $\mathsf{Show}$ proof with a
simulation for its true canonical statement
$x=(\rho,z_{\mathsf{root}},z_{\mathsf{pre}},\tau_{\mathsf{pre}})$.
The handles, signing calls, ledger entries, claimed attributes, and
public context checks are unchanged. Soundness of the handle checks and
honest credential evolution ensures the simulated statements are true.
Adaptive auxiliary-input zero knowledge bounds this change even when the
adversary has real credential tokens and other oracle outputs. The same
canonical statement is recorded in the show log in both games, so a
change of proof bytes is not a fresh statement.

On a fresh accepting presentation, use extraction in the simulated
experiment to obtain a finite trace with one issue, zero or more
delegations, and one presentation, for the same trusted root, public
context, and disclosed authorization $A^*$. The extractor is tied
to the accepting output and the signing transcript. The predicate at
the end opens the predecessor commitment, extracts its signing-key
witness, and enforces the authorization restriction. When adjacent
witnesses give different openings of the same commitment, the reduction
can output a binding violation. Canonical context and attribute encodings
and the presentation binding ensure that the inspection concerns the
same accepted public statement.

Scan the trace from the root, matching each signed transition to the
ledger entry for the signing key. Initially the signer is the honest root.
While the signer is honest, if its extracted signed digest is absent
from its signing ledger, the trace contains an EUF-CMA forgery.
If the digest is present, compare the structured message and openings
with the ledger record. A different structured message with that digest
gives a collision in $H_{\mathsf{sig}}$, and different openings of the same
state commitment give a binding violation. Otherwise the entry fixes the
matched transition and its registered destination, so the scan can
continue from this state. The comparison is of signed messages, not
signature bytes.

If a matched transition first reaches an adversary-controlled holder,
the corresponding authorization is in the authorized closure. This
includes credentials independently obtained by holders before a later
corruption, as well as those exposed or delivered thereafter. Every
later delegation and the final presentation preserve the authorization
restriction, so transitivity puts $A^*$ in the closure, contrary to the
winning condition. Thus the scan needs no ledger entries for transitions
performed after reaching an adversarial key.

If there is no missing signature and no binding violation, the only
remaining case is that the trace reaches an uncorrupted, registered
honest holder. The extracted terminal witness permits a signature on a
digest that has not been signed for that holder
(Definition~\ref{def:pcc-key-relation}), again yielding EUF-CMA. This
case is essential: merely identifying a fully authorized lineage does
not show that the adversary could present as its terminal holder.

For each signature case the reduction guesses the implicated honestly
generated verification key (including the root for issuance) and embeds its
EUF-CMA challenge at that registration. It uses the challenger for that
key's authorized signing calls and aborts if that key is corrupted.
Credential proofs use public keys, state openings, and the already obtained
signatures; the guessed key's secret is not needed to generate them.
For honest showings under the guessed key, the reduction already uses
simulated proofs for true statements, so it does not need this secret
until extraction. Its guess succeeds with inverse polynomial
probability, because the implicated key is still honest. Signing after
terminal extraction is performed with the extracted witness, not with the
EUF-CMA oracle. Extraction cannot add queries to the signing ledger or
rewind the external signer. For $q_{\mathsf{key}}$ candidate keys the
bound is of the form
\[
\begin{aligned}
\mathsf{Adv}^{\mathsf{uf}}(\mathcal A)
\le{}&
 \epsilon_{\mathsf{zk}}+\epsilon_{\mathsf{ext}}
 +\epsilon_{\mathsf{bind}}\\
 &+q_{\mathsf{key}}
   \mathsf{Adv}^{\mathsf{euf}}_{\Sigma}(\mathcal B)
 +\negl(\lambda).
\end{aligned}
\]
Here $\epsilon_{\mathsf{bind}}$ covers commitment and canonical hashed
statement binding, and the negligible term includes honest proof
generation failures. The argument concerns the public authorization of
a fresh presentation. In this construction the public presentation has
no token-specific identifier on which to base a separate ownership test.
\end{proof}

\subsection{Enhanced Delegation Anonymity}
\label{app:pqdac-eda}

\begin{lemma}[Verifier-side anonymity]
\label{lem:pqdac-verifier-anon}
If the presentation proofs satisfy the zero-knowledge interface of
Definition~\ref{def:pcc-proof-interface} and all non-proof parts of the
observable presentation are determined by its public statement and
allowed auxiliary information, then PCC satisfies
Definition~\ref{def_verif_Anony}.
\end{lemma}

\begin{proof}
For either admissible path, the presentation proof has the same statement
\[
 x_{\mathsf{pre}}=
 (\rho,z_{\mathsf{root}},H_{\mathsf{pre}}(\mathsf{pre},ctx,A_{\mathsf{pre}}),
 \tau_{\mathsf{pre}}(ctx,A_{\mathsf{pre}})).
\]
The verifier does not receive the predecessor commitment, predecessor
proof, or terminal key in the native presentation. The two worlds are
indistinguishable from the same experiment that simulates the fresh proof
using only $x_{\mathsf{pre}}$ in the common setup. Auxiliary inputs
permitted by the game, including known proof objects, remain available
in this hybrid. Any extra application fields must be matched as required
by the game; no pseudonym is added by the PCC wrapper.
\end{proof}

\begin{lemma}[Holder-side topology obliviousness]
\label{lem:pqdac-holder-obliv}
If credential proofs satisfy Definition~\ref{def:pcc-proof-interface}
and the token and proving interfaces expose no extra topology-dependent
metadata, then PCC satisfies Definition~\ref{def_holder_Anony}.
\end{lemma}

\begin{proof}
The challenge fixes the same terminal key pair, authorization, public
parameters, presentation inputs, and allowed incoming-edge information.
Couple the honest final handoffs by using the same freshly sampled
opening randomness $r_H$ in both experiments; each marginal has the
prescribed honest distribution. The terminal commitment
$z_H=\commit(\upk_H,A_H;r_H)$ is then common. The honestly delivered
token contains $(\rho,z_H,\pi_H,\upk_H,A_H,r_H)$, hence its only field
still dependent on the choice of remote path is the credential proof
$\pi_H$ and any metadata attached to it. The proof may nevertheless
be correlated with auxiliary information about the prefixes, so we use
auxiliary-input zero knowledge for an honestly generated handoff,
at the common statement
$x_H=(\rho,z_{\mathsf{root}},z_H,\tau_{\mathsf{cred}})$.

Replace only $\pi_H$ by a simulation for $x_H$. The holder's own
presentation is then obtained by actually running the prescribed
$\pccderive$ on the simulated accepting predecessor, the same local
opening and secret, and fresh holder presentation randomness. This is
efficient post-processing in the common setup and oracle environment:
the wrapper needs no predecessor witness beyond the delivered proof.
It preserves the correlations among the credential bytes, the actual
inputs to the holder's proving invocation, the holder's coins if
observable, and the resulting presentation. In particular, we do not
simulate the presentation independently of its observable proving inputs.
Both hybrid views now depend only on the common terminal data, allowed
auxiliary information, and $x_H$, so they are indistinguishable.
This does not hide credentials or proof-generation traces that a
colluding predecessor independently discloses. Any length,
verifier-data identifier, parser field, or other exposed metadata that
varies with remote depth would violate the stated interface condition.
\end{proof}

\begin{proof}[Proof of Theorem~\ref{thm:pqdac-eda}]
These two lemmas establish the two privacy experiments in the stated
leakage and oracle model.
\end{proof}

\subsection{Post-Quantum Security in the QROM}
\label{app:pqdac-qrom}
\begin{proof}[Proof of Theorem~\ref{thm:pqdac-qrom}]
Assume the same interface, binding, usable signing witness, and
signature-security hypotheses against quantum polynomial-time
adversaries in the common model of Appendix~\ref{app:qrom-model}.
The integrity reduction uses the extractor for the observed accepting
output and does not implement a classical rewind of an external quantum
adversary or signer. The anonymity reductions use computational
indistinguishability and the honest presentation operation as
post-processing. Thus the same reductions apply with the
corresponding quantum advantages. This is a statement about preservation
of the assumed properties, not an instantiation proof for the measured
FRI and Poseidon parameters.
\end{proof}

\section{Wrapper-Level Security Games for PCC}
\label{app:pcc-security}
The wrapper forwards recursive steps to $\nivcprove$. Its security
interface concerns the recursive public statement
$x=(\rho,z_{\mathsf{root}},z_{\mathsf{out}},\tau_{\mathsf{pub}})$, distinct
from the holder's full token or the prover's private inputs. We use the
typed terminal conditions $\tau_{\mathsf{cred}}=(\mathsf{tag}=\mathsf{cred})$
and $\tau_{\mathsf{pre}}(ctx,A_{\mathsf{pre}})
=(\mathsf{tag}=\mathsf{pre},ctx,A_{\mathsf{pre}})$.

\begin{definition}[Valid PCC lineage]
\label{def:valid-lineage}
A lineage for $x$ is a finite list of states
$(s_0,\ldots,s_n)$, selectors $(j_1,\ldots,j_n)$ and witnesses
$(w_1,\ldots,w_n)$. It starts with
$s_0=z_{\mathsf{root}}=H_{\mathsf{state}}(\mathsf{tag}_{\mathsf{issue}},\rho)$,
ends with $s_n=z_{\mathsf{out}}$, and every step verifies
$\phi_{j_i}(\rho,s_{i-1},s_i,w_i)=1$ under the same $\rho$.
For a credential terminal condition the selectors have the form
$\mathsf{issue}\,\mathsf{del}^{*}$; for a presentation terminal condition
they have the form $\mathsf{issue}\,\mathsf{del}^{*}\,\mathsf{pre}$.
In the latter case, the presentation witness uses the public
$(ctx,A_{\mathsf{pre}})$ and the last state is
$H_{\mathsf{pre}}(\mathsf{pre},ctx,A_{\mathsf{pre}})$.
At each non-base step the prior proof is checked for the credential
terminal condition and the same predecessor state. The depth is bounded
by the experiment's proof horizon, not exposed to the verifier.
\end{definition}

\begin{definition}[Wrapper-level completeness]
\label{def:completeness}
An honestly invoked $\pccderive^*$ accepts its valid base or recursive
inputs and yields an accepting proof for the corresponding typed
statement, except with negligible probability. The base case uses only
an issue step from $z_{\mathsf{root}}$ with predecessor $\bot$. The other
steps extend a credential proof by delegation or close it by presentation.
\end{definition}

\begin{definition}[Wrapper-level simulation extractability]
\label{def:sim_ext}
Consider the wrapper experiment with access to honest credential
generation and simulated proofs for adaptively chosen true
statements. Log each simulated canonical public statement $x$ in
$Q_{\mathsf{sim}}$. The probability of an accepting output
$(x^*,\pi^*)$ with $x^*\notin Q_{\mathsf{sim}}$ for which the extractor
does not return a valid lineage for that output is negligible. The
extractor and auxiliary-oracle restrictions are those of
Definition~\ref{def:pcc-proof-interface}. The log is over statements,
not over proof encodings; in particular, the premise is not a claim
about an altered proof of an already simulated statement.
\end{definition}

\begin{definition}[Wrapper-level zero knowledge]
\label{def:zk}
In the real experiment a query supplies a valid wrapper proving input
for $x$ and receives the honest proof. In the simulation experiment the
same admissibility checks are applied, and the proof is simulated using
only $x$ and the simulator's state. Invalid queries return $\bot$ in both
experiments. No efficient adversary distinguishes the experiments,
including the permitted auxiliary input and system responses, with
non-negligible advantage. The simulated proof is returned in the same
observable interface and layout as the real proof; private mode and
predecessor-state data are not simulator inputs.
\end{definition}

\begin{theorem}[Security of the PCC wrapper]
\label{secpcc}
For a PCC-compatible NIVC satisfying
Definition~\ref{def:pcc-proof-interface} for the typed execution
language, $\pccderive^*$ and the corresponding verifier preserve
wrapper-level completeness, simulation extractability, and zero
knowledge.
\end{theorem}

\begin{proof}
A valid wrapper input supplies the selected transition witness and the
prior credential proof, or the designated empty proof for issuance. It
invokes the same honest NIVC prover on that input, so completeness is
preserved. The wrapper verifier checks the public root, terminal type
and presentation binding, then invokes the NIVC verifier on exactly
$x$. The extraction experiment forwards the same true statements to
the backend simulator and logs those statements. Any fresh wrapper
acceptance is therefore a fresh backend acceptance for the same $x$,
and the returned typed lineage verifies all PCC steps. For zero
knowledge, replace the backend answers by its simulations of these
same statements, keeping the wrapper's public checks and deterministic
packaging. Under the auxiliary-input and layout hypotheses this preserves
the adversary's view. The wrapper does not derive typed recursive
extraction or auxiliary-input zero knowledge from plain soundness.
\end{proof}

\end{document}